\documentclass[sigplan,nonacm]{acmart}
\usepackage{tikz}
\usepackage{amsmath}
\usepackage{enumitem}
\usepackage{kotex}
\usepackage[linesnumbered,ruled,vlined]{algorithm2e} 
\usepackage{subfig}
\usepackage{graphicx}
\usepackage{multirow}

\usepackage{adjustbox}

\SetKwInput{Input}{Input}
\SetKwInput{Output}{Output}
\SetKwProg{Fn}{Function}{:}{}
\SetKwFunction{MergeCentroids}{MergeCentroids}
\SetKwFunction{FilteringCentroids}{FilteringCentroids}
\SetKwFunction{AdjustCentroids}{AdjustCentroids}
\SetKwFunction{UpdateCentroids}{UpdateCentroids}
\SetKwFunction{FindClosestCentroid}{FindClosestCentroid}
\SetKwFunction{CosineSimilarity}{CosineSimilarity}
\SetKwFunction{TotalMemoryUsage}{TotalMemoryUsage}

\newcommand{\eg}{e.g.,}

\usepackage{dblfloatfix}
\usepackage[normalem]{ulem}

\definecolor{forestgreen}{rgb}{0.0, 0.5, 0.0}

\usepackage{booktabs} 
\usepackage{caption}

\definecolor{darkblue}{rgb}{0.0, 0.0, 0.54}

\newcommand{\ours}{SISO}

\newcommand{\SEC}[1]{$\S$\ref{#1}}
\newcommand{\FIG}[1]{Fig.~\ref{#1}}

\newcommand{\TAB}[1]{Table~\ref{#1}}
\newcommand{\EQ}[1]{Eq.~\ref{#1}}
\newcommand{\OBS}[1]{Observation~\#\ref{#1}}

\usepackage{tikz}

\newcommand{\wcircled}[1]{
    \setbox0=\hbox{#1}%
    \dimen0\wd0%
    \divide\dimen0 by 2%
    \begin{tikzpicture}[baseline=(a.base)]%
        \useasboundingbox (-\the\dimen0,0pt) rectangle (\the\dimen0,1pt);
        \node[color=black!50!black,circle,draw,outer sep=0pt,inner sep=0.05ex] (a) {#1};
    \end{tikzpicture}
}

\newcommand{\bcircled}[1]{
    \setbox0=\hbox{#1}%
    \dimen0\wd0%
    \divide\dimen0 by 2%
    \begin{tikzpicture}[baseline=(a.base)]%
        \useasboundingbox (-\the\dimen0,0pt) rectangle (\the\dimen0,1pt);
		\node[fill=black!100,circle,text=white,outer sep=0pt,inner sep=0.05ex] (a) {#1};
    \end{tikzpicture}
}

\newcommand{\redcircled}[1]{
    \setbox0=\hbox{#1}%
    \dimen0\wd0%
    \divide\dimen0 by 2%
    \begin{tikzpicture}[baseline=(a.base)]%
        \useasboundingbox (-\the\dimen0,0pt) rectangle (\the\dimen0,1pt);
		\node[fill=red!100,circle,text=white,outer sep=0pt,inner sep=0.05ex] (a) {#1};
    \end{tikzpicture}
}

\newcommand{\bluecircled}[1]{
    \setbox0=\hbox{#1}%
    \dimen0\wd0%
    \divide\dimen0 by 2%
    \begin{tikzpicture}[baseline=(a.base)]%
        \useasboundingbox (-\the\dimen0,0pt) rectangle (\the\dimen0,1pt);
		\node[fill=blue!100,circle,text=white,outer sep=0pt,inner sep=0.05ex] (a) {#1};
    \end{tikzpicture}
}

\usepackage{comment}

\newcounter{take}
\newcommand{\head}[1]{{\noindent\textbf{#1.}\xspace}}
\usepackage[most]{tcolorbox}

\newcommand{\lesson}[1]{%
  \par\addvspace{0pt}
  \refstepcounter{take}%
  \begin{tcolorbox}[enhanced,
    before skip=4pt,
    breakable,                    
    colback=lightgray!35!white,   
    boxrule=0.5pt,                  
    arc=2pt,                      
    after skip=4pt,
    left=0pt,right=0pt,top=0.1pt,bottom=0pt 
  ]%
    \head{Observation \#\thetake}\space{#1}%
  \end{tcolorbox}%
  \par\addvspace{0pt}
  \noindent\ignorespaces
}

\begin{document}



\title[Rethinking Caching for LLM Serving Systems: Beyond Traditional Heuristics]{Rethinking Caching for LLM Serving Systems:\\Beyond Traditional Heuristics}


\author{Jungwoo Kim}
\email{jungwoo@dgist.ac.kr}
\affiliation{%
  \institution{DGIST}
  \country{South Korea}
}

\author{Minsang Kim}
\email{kimmsang96@dgist.ac.kr}
\affiliation{%
  \institution{DGIST}
  \country{South Korea}
}

\author{Jaeheon Lee}
\email{jaeheon@postech.ac.kr}
\affiliation{%
  \institution{POSTECH}
  \country{South Korea}
}

\author{Chanwoo Moon}
\email{ixora99@dgist.ac.kr}
\affiliation{%
  \institution{DGIST}
  \country{South Korea}
}

\author{Heejin Kim}
\email{noah211@postech.ac.kr}
\affiliation{%
  \institution{POSTECH}
  \country{South Korea}
}

\author{Taeho Hwang}
\email{taeho.hwang@sk.com}
\affiliation{%
  \institution{SK hynix}
  \country{South Korea}
}

\author{Woosuk Chung}
\email{woosuk.chung@sk.com}
\affiliation{%
  \institution{SK hynix}
  \country{South Korea}
}

\author{Yeseong Kim}
\email{yeseongkim@dgist.ac.kr}
\affiliation{%
  \institution{DGIST}
  \country{South Korea}
}

\author{Sungjin Lee}
\email{sungjin.lee@postech.ac.kr}
\affiliation{%
  \institution{POSTECH}
  \country{South Korea}
}

\begin{abstract}
Serving Large Language Models (LLMs) at scale requires meeting strict Service Level Objectives (SLOs)
under severe computational and memory constraints. Nevertheless, traditional caching strategies fall short:
exact-matching and prefix caches neglect query semantics, while state-of-the-art semantic caches remain confined to 
traditional intuitions, offering little conceptual departure. Building on this, we present SISO, a semantic caching system
that redefines efficiency for LLM serving. SISO introduces centroid-based caching to maximize coverage with minimal memory, locality-aware replacement to preserve high-value entries, and dynamic thresholding to balance accuracy and latency under varying workloads. 
Across diverse datasets,
SISO delivers up to 1.71× higher hit ratios and consistently stronger SLO attainment compared to state-of-the-art systems.
\end{abstract}

\maketitle


%

\section{Introduction}
The rise of Large Language Models (LLMs) has triggered a paradigm shift in various 
applications~\cite{gpt3}, transforming dominant 
information retrieval from traditional data management systems (e.g., databases and data warehouses) to LLM inference.
This transformation, however, incurs
substantially higher energy consumption and computational demands.
As a result, there is a crucial need to develop efficient LLM serving (or inference) systems that ensure high-quality user experience by meeting strict Service Level Objectives (SLOs) at lower energy usage and operational costs.
Toward this goal, many have explored various optimization strategies,
including LLM  task
scheduling~\cite{orca, alpaserve}, GPU memory optimization~\cite{flexgen, vllm}, and model scaling~\cite{tabi, proteus}.

Despite their effectiveness, 
existing approaches often require system-wide modifications, limiting their portability across various platforms and their compatibility with other techniques.
As a complementary direction, caching can be employed to reuse previously computed outputs -- 
an idea long established in traditional applications like SQL query caching~\cite{sqlcaching,sqlcaching2,sqlcaching3}.
Current LLM caching solutions, however, are built upon conventional intuition for caching,
treating LLM queries as structured strings or instructions like SQL queries~\cite{langchain, azure-openai, mooncake, cacheblend}. 
This prevents us from fully exploiting the potential of
LLM caching.
For example, ``What is semantic caching?'' and ``Explain semantic caching'' are semantically similar and 
expected to produce similar outputs, but existing LLM caching fails to recognize them as the same query.

In light of these limitations, semantic caching has recently emerged, focusing on the meaning of queries.
This approach offers promising opportunities for extending
caching methodologies to LLMs by incorporating semantic similarity into query caching.
However, prior work on semantic caching has mostly been studied in the context of LLMs, such as similarity matching algorithms~\cite{gptcache, meancache, cacheme}, with limited system-level design considerations.
For example, state-of-the-art (SOTA) systems (e.g., GPTCache~\cite{gptcache}) naively adopt traditional policies when promoting or evicting queries, without tailoring them to the unique characteristics of LLM workloads.
This oversight leads to several limitations.

First, existing semantic caching treats individual queries as a unit of caching, promoting or demoting them independently.
Although it is a common practice in conventional caching,
this approach leads to storing duplicate queries with nearly identical meanings, resulting in a waste of cache space.
Second, existing semantic caching relies on recency or frequency for eviction, which is invoked frequently when the cache 
needs to free up space for new queries.
While it seems reasonable, 
such replacement policies often make wrong decisions in LLM workloads,
evicting valuable queries and thereby reducing overall hit ratios.
Third, existing semantic caching requires all requests to pass through the semantic caching layer, regardless of whether caching is beneficial.
Unlike traditional caching that always returns accurate responses, semantic caching returns similar outputs for input queries, inherently trading accuracy for performance.
Thus, semantic caching should be applied selectively, when performance has a higher priority than accuracy (e.g., when heavy computation might lead to the violation of SLOs).

In this paper, we propose \ours{} 
(\textbf{S}imilar \textbf{I}nput? \textbf{S}imilar \textbf{O}utput!), 
a novel semantic caching system that moves beyond traditional heuristics, optimized for the semantic characteristics of LLM queries and workloads.
\ours{} incorporates three features. The first is \textit{centroid-based caching}. 
Instead of caching individual queries, \ours{} only stores centroids 
which represent many semantically similar queries.
By keeping only valuable queries, \ours{} uses the cache space efficiently without any redundant caching.

The second is \textit{semantic locality-aware centroid replacement}.
Here, semantic locality refers to the degree to which a centroid can represent a broad range of semantically similar queries.
\ours{} prioritizes caching centroids with strong semantic locality, while evicting those with weaker semantic locality.
Since semantic locality remains relatively stable over time, replacements are triggered occasionally by monitoring long-term query behaviors. 
This approach yields higher cache hit ratios than traditional LRU and LFU policies.

The third is \textit{dynamic threshold adjustment} to balance output quality and serving latency. 
While serving input queries, \ours{} dynamically adjusts a similarity threshold, which decides whether input queries are sufficiently similar to cached centroids or not.
When workloads are intensive, the threshold can be relaxed to increase cache hits, alleviating pressure on the LLM serving system. 
Conversely, when workloads are light, it can be tightened to maximize response quality.

To evaluate the effectiveness of \ours{}, we conduct experiments using various real-world datasets~\cite{qqp,mqp,mrpc,reddit} with two LLMs: LLaMa-3.1 8B and 70B~\cite{llama3}. 
For evaluation, we compare \ours{} with the SOTA LLM serving system, vLLM~\cite{vllm}, and semantic caching system, GPTCache~\cite{gptcache}.
Through extensive experiments, we obtain the following key results. 
First, by caching centroids rather than individual outputs, \ours{} can efficiently utilize available cache space, thereby exhibiting 1.54$\times$ higher hit ratios than GPTCache on average. 
Second, through semantic locality-aware replacement, \ours{} improves the average cache hit ratios by 1.71$\times$ 
by maintaining
more valuable centroids in the cache. 
Finally, by dynamically adjusting its similarity threshold, \ours{} achieves higher SLO attainment under intensive and/or highly variable workloads, surpassing both vLLM and GPTCache.
This improvement comes with only a marginal accuracy drop of 6.9\%.
Under light workloads, \ours{} maintains the same accuracy as vLLM without any loss of output quality.

While \ours{} beats existing systems, 
its benefits are limited to single-turn queries for
specific types of tasks, such as information and advice seeking. 
The current version cannot support context-dependent tasks (i.e., multi-turn queries) and performs relatively poorly on coding and debugging tasks.
However, given single-turn queries dominate user interactions (99\% in API calls and 67\% in chatbots~\cite{cachewild}), and that information- and advice-seeking 
account for 53.1\% of all queries, these limitations do not negate \ours{}'s value but highlight a clear need for future research in semantic caching.

This paper is organized as follows: \SEC{sec:background} gives the background and related work.
After presenting key design principles of \ours{} in \SEC{sec:desprin}, we detail the implementation of \ours{} in \SEC{sec:implementation}, explaining how \ours{} is implemented based on its design principles. 
After showing experimental results in \SEC{sec:evaluation},
we discuss limitations of \ours{} and directions for further improvements 
in \SEC{sec:discussion}. We finally conclude in \SEC{sec:conclusion}.
\section{Background and Related Work}
\label{sec:background}

This section gives an overview of LLM serving systems, along with reviews of prior efforts to improve their performance (\SEC{sec:serving} -- \SEC{sec:cachellmsv}).

\subsection{LLM Serving Systems}
\label{sec:serving}
The rapid advent of LLMs has increased the demand for developing LLM serving systems, which are designed to deploy LLMs for real-time inference tasks~\cite{tensorrt-llm, vllm, deepspeed-fastgen}. 
The primary challenge in building these systems lies in efficiently managing computational resources (e.g., GPUs and CPUs) to consistently meet strict latency requirements,
SLOs~\cite{sarathi-serve}.

The difficulty in meeting SLOs originates from the LLM inference process, which is divided into two distinct phases. 
The first is the compute-intensive prefill phase, where the model processes the user's entire query to generate the first token. 
This initial burst of computation determines the Time-to-First-Token (TTFT), which governs the user's perception of initial responsiveness. 
The second is the decoding phase that sequentially generates the remaining response one token at a time. 
The latency of each step is measured as Time-between-Tokens (TBT). 
The decoding has relatively low computation, but since it relies on a huge KV cache of intermediate attention states to avoid costly recomputation, it is considered memory intensive.
Consequently, these two metrics, TTFT and TBT, are used to define SLOs, 
which are typically set to match human reading speed (e.g., $\sim$50ms per token) or 
constrained to less than 1.3$\times$ the latency observed under a zero-load setup~\cite{aladdin, sloscheduling}.

LLM serving systems should be designed to avoid violating SLOs not only during periods of low demand, but also during peak demand.
The brute-force solution to
achieve this is resource over-provisioning, a practice of allocating more computing resources than required during regular use~\cite{nexus}.
Nevertheless, over-provisioning is a costly safeguard -- wasteful for most of the time, yet indispensable to prevent SLO violations during the peak. As a result, excess resources remain idle during off-peak hours, inevitably increasing hardware acquisition and operational costs.
This inefficiency highlights the need for more cost-efficient serving strategies.

\subsection{Optimization of LLM Serving Systems}
\label{sec:optllmsv}
There have been numerous efforts to improve SLO compliance and cost efficiency~\cite{sola,spotserve,alphaserve}.
These studies have evolved along two primary axes: (i) computational optimization to maximize GPU throughput
and (ii) memory optimization to support a higher volume of concurrent requests.

Many have addressed computational inefficiency by 
optimizing GPU utilization and enhancing pipeline efficiency across
prefill and decoding phases with different resource demands.
Common techniques include task scheduling tailored to LLM~\cite{orca, sarathi-serve, distserve}, GPU kernel optimizations~\cite{flashdecoding,flashinfer, flashattetnion} and system-accelerator co-design~\cite{aladdin, neupim, powerinfer}. These reduce pipeline bubbles and improve throughput, but demand fundamental modifications to existing systems.

Some have attempted to optimize memory utilization by 
improving memory management or reducing memory footprints.
Prior studies developed specialized custom attention kernels to manage KV cache in non-contiguous blocks without memory fragmentation,
while some introduced a contiguous virtual address for efficient KV cache management~\cite{vllm, vattention, memserve, prism, actcheckpoint}.
Other methods explored footprint reduction through model/KV cache quantization requiring specific tensor layouts or custom GPU kernels~\cite{quan1, kvquant, proteus}, and offloading the KV cache to host DRAM or even SSD~\cite{flexgen, infinigen, flashgen}.

The above approaches improve computation and memory efficiency. However, their reliance on architecture-level modifications and custom kernels is disruptive to the serving stack, making them difficult to deploy and even harder to integrate with one another.
Such incompatibility significantly undermines their practicality in real-world deployments.

\subsection{In-memory Caching for LLM Serving}
\label{sec:cachellmsv}
The aforementioned limitations have motivated the 
need for a non-intrusive complementary solution.
One proven approach from traditional database systems is caching, where in-memory stores~\cite{redis, memcached} have improved RDBMS performance without changes to the underlying logic. Analogously, caching offers a promising approach for improving LLM serving efficiency while preserving system compatibility.

The most natural extension to LLM serving is \textit{exact-match caching}: store a response and serve it again for identical inputs.
Due to its simplicity, exact-match caching serves as a basic function in many LLM serving systems~\cite{langchain, azure-openai, mooncake, cacheblend}.
However, its practical impact is limited, as even trivial variations in prompts lead to cache misses.

A more LLM-aware variant is \textit{prefix caching}, which exploits
the auto-regressive property of LLMs~\cite{allyouneed}.
For instance, the prompts ``Summarize the plot of Star Wars'' and ``Summarize the plot of Star Trek'' share the prefix (i.e., first five tokens).
By reusing the KV states of the shared prefix, the system can bypass redundant computations for the latter query.
Prior studies have improved its effectiveness by extending cache capacity from DRAM to SSD~\cite{mooncake}, developing faster overlap detection~\cite{sglang}, and allowing slight differences in prefix~\cite{cacheblend}.
Still, prefix caching relies on token-level overlap; in practice, many hits come from system prompts rather than user content~\cite{cachewild}, and paraphrases with different tokens remain as misses.

As an alternative to exact-match and prefix caching, 
\textit{semantic caching} has gained attention, which reuses query outputs for new inputs that are semantically similar rather than textually identical.
GPTCache~\cite{gptcache} is a representative open-source implementation of semantic caching.
It keeps a vector representation of an input query along with its corresponding output in memory.
When a new request arrives, the semantic caching system computes a vector representation of the input and compares it with cached vectors to identify a semantically similar one, returning its output 
without LLM execution.
Subsequent work has sought to raise hit rates and retrieval efficiency via knowledge distillation~\cite{cacheme}, advanced pattern matching~\cite{scalm}, and federated learning~\cite{meancache}.

Despite its potential, SOTA semantic caching systems remain limited: they blindly apply traditional caching heuristics that are not suited for LLMs.
First, existing approaches cache individual query embeddings. Due to this granularity, however, even nearly identical queries may be stored multiple times, resulting in wasted cache space.

Secondly, cache eviction is often tied to generic replacement heuristics, such as recency or frequency, that are triggered when the cache is full. 
While such a short-window replacement suffices in traditional caches 
where reuse is usually bursty and immediate, 
LLM serving demands a longer temporal horizon. Queries that appear infrequently or with longer gaps can still be semantically related to future requests. Thus, prematurely evicting them sacrifices reuse opportunities that are especially valuable in semantic caching.

Lastly, the current system always uses cached outputs in the case of a hit, regardless of whether the system is idle or busy. Unlike traditional caches, semantic caching inherently involves approximation -- the returned response may only be ``similar'' to the true one. This approximation is only worthwhile when its performance benefits outweigh the degradation in output quality. 
Nevertheless, existing semantic caching still relies on fixed cache hit mechanisms, which may not always align with practical serving conditions.
\section{Design Principles of \ours{}}
\label{sec:desprin}
To overcome the limitations of caching for the LLM serving system, we propose \textit{\ours{}}, a novel semantic caching framework. \ours{} is designed to leverage the semantic nature of LLM queries while maintaining the practical benefits of traditional caching, such as simple deployment and no architectural changes to underlying serving systems.
The proposed \ours{} manages cache space by considering the unique properties of vectors and analyzing long-term trends of input queries.
Additionally, it dynamically adjusts its policies in response to the status of the serving system.
In this manner, it enables fast LLM serving performance and delivers the highest possible accuracy with better SLO compliance.

\begin{figure}
    \centering
    \includegraphics[width=0.90\linewidth]{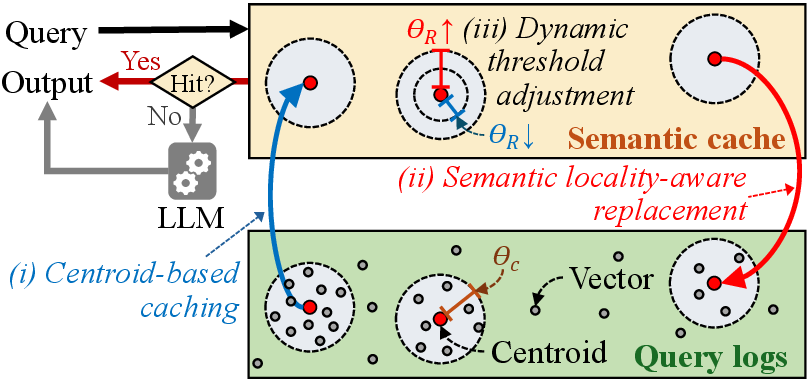}
    \caption{Overall organization of \ours{}}
    \Description{Overall organization of \ours{}}
    \label{fig:overview}
\end{figure}

The overall organization of \ours{} is illustrated in~\FIG{fig:overview}, which is built upon three unique design principles:
(\textit{i}) \textit{centroid-based caching}, (\textit{ii}) \textit{semantic locality-aware centroid replacement}, and (\textit{iii}) \textit{dynamic threshold adjustment}.

\textbf{Centroid-based caching}.
Instead of storing vectors of individual queries, \ours{} caches only centroid vectors that represent groups of similar queries, allowing more efficient use of the cache space. 
To construct these centroids, \ours{} analyzes query logs (typically recorded by LLM serving systems~\cite{wildchat}) containing queries and their responses, clusters the queries based on semantic similarity, and inserts the resulting cluster centroids into the cache.
This approach enables efficient use of limited memory without sacrificing output accuracy (see~\SEC{sec:centbasedcaching}).

\textbf{Semantic locality-aware centroid replacement}.
Unlike existing semantic caching methods that evict vectors based on short-term recency or frequency~\cite{gptcache}, \ours{} exploits a different type of locality, \textit{semantic locality}, which can be identified by observing long-term behaviors of input queries.
Semantic locality refers to the property that centroids with a wide similarity coverage, representing many other vectors, receive the majority of queries. 
Since the popularity of centroid evolves slowly, \ours{} replaces the centroids exhibiting a low semantic locality with higher-locality ones 
through re-clustering over a long-term time window (see~\SEC{sec:semwarerp}).

\textbf{Dynamic threshold adjustment}.
In contrast to existing semantic caching that uses a fixed similarity threshold $\theta_{R}$ for retrieval~\cite{gptcache}, \ours{} adjusts the threshold according to the intensity of workloads.
We observe that output quality is linearly proportional to the similarity threshold, meaning that adjusting the threshold directly controls the quality of responses.
Leveraging this property, \ours{} dynamically balances accuracy and performance in a workload-aware manner.
Under heavy workloads, it lowers the threshold to increase the chances of finding similar vectors in the cache, thereby raising hit ratios and reducing the load on the underlying LLM serving system.
Conversely, under a light workload, it increases the threshold or turns off the semantic cache to maximize the quality of responses (see~\SEC{sec:adjsimths}).

We now present our analyses and five key observations that support the fundamental design principles for \ours{}.

\subsection{Centroid-based Caching}
\label{sec:centbasedcaching}

To show the advantages of centroid-based caching over vector-level caching, we analyze real-world datasets to assess the accuracy of caching centroids and its memory efficiency.

We use Quora Question Pair (QQP), Microsoft Research Paraphrase Corpus (MRPC), and Medical Question Pairs (MQP)~\cite{qqp, mrpc, mqp}, which are used to develop algorithms that identify duplicate questions or sentences that have the same meaning.
Each dataset item is a set of \{text1, text2, is\_duplicate\}, where text1 and text2 are English texts and is\_duplicate is a binary label indicating whether the two texts are duplicates.
We measure the cosine similarity of two types of text pairs: one with duplicate text pairs and another with non-duplicate text pairs.
To compute the similarity, we utilize a pre-trained embedding model, \texttt{paraphrase-albert- small-v2}~\cite{sentence-transformers} (see~\SEC{sec:sisocluster} for more details on our choice).

\begin{figure}[t]
    \centering
        \begin{minipage}[t]{0.22\textwidth}
    	\centering
	        \includegraphics[origin=c,width=\linewidth]{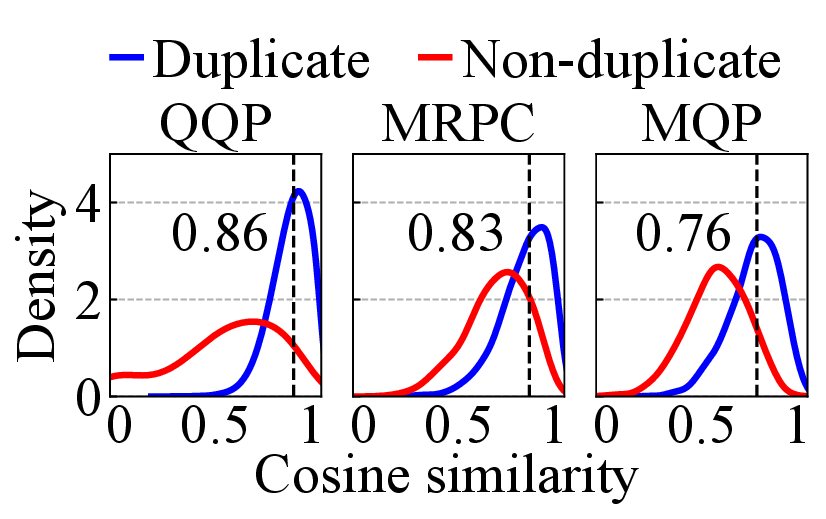}
            \caption{Cosine similarity analysis}
	    \label{fig:dupnondup}
        \end{minipage}
        \hspace{0.1pt}
        \begin{minipage}[t]{0.243\textwidth}
        \centering
            \includegraphics[origin=c,width=\linewidth]{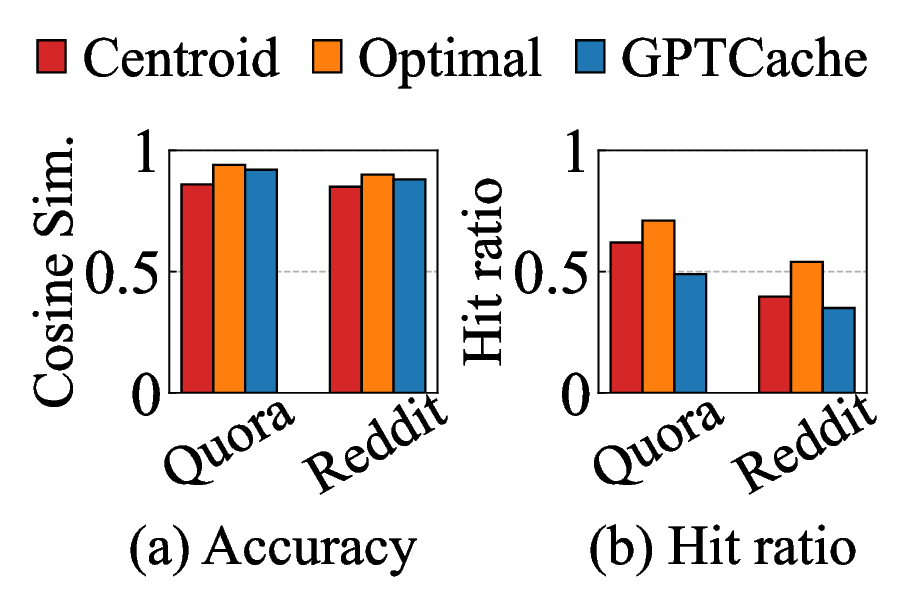}
            \caption{Impact of centroid-based caching}
            \label{fig:centvsopt}
        \end{minipage}
\end{figure}

\FIG{fig:dupnondup} plots the PDF graphs of cosine similarities obtained from duplicate (a blue line) and non-duplicate (a red line) pairs. 
Duplicate pairs have a high median cosine similarity, averaging 0.82 across datasets.
In contrast, non-duplicate pairs exhibit a low median cosine similarity of 0.62 on average.
These results show that when threshold values are set sufficiently high, for example, higher than the median values, 0.86 for QQP, 0.83 for MRPC, and 0.76 for MQP, two texts are highly likely to convey the same meaning.

\lesson{A group of texts with high cosine similarities are likely to have the shared meaning, and thus are considered duplicates.}
\label{obs:observation1}

\textbf{Impact on accuracy.}
Based on the above observations, we perform an analysis using realistic datasets to understand how caching only centroids affects accuracy.
Here, a centroid is a vector that represents the central or average point of a group of vectors that have similar
cosine similarity.

We use two datasets: 600K questions collected from Quora and 600K questions obtained from the Reddit dataset~\cite{reddit}. 
The dataset is divided into 95\% for training and 5\% for testing.
Using LLaMa-3.1-8B, we generate answers for questions in the training dataset, creating pairs of <question, answer>.  
We then perform clustering based on cosine similarity and select a centroid to represent each cluster.
We set the clustering threshold $\theta_{C}$ to 0.86 for both datasets, a value chosen to be sufficiently high based on \OBS{obs:observation1}.
The community detection algorithm~\cite{comdet} is used for clustering (see~\SEC{sec:sisocluster} for more details), and we found 60K centroids in each dataset.

We compare three systems: \textsf{Centroid}, which caches 60K centroids (ours); \textsf{GPTCache}, which uses an LRU policy with a 60K vector limit; and \textsf{Optimal}, which retains all 600K vectors in memory and serves as an impractical oracle due to its high memory consumption.
As with the training dataset, we create <question, answer> pairs for the test dataset.
We replay the collected dataset by sending a sequence of the same questions to both systems based on their timestamps. 
The similarity threshold for retrieval, $\theta_{R}$ (which judges a cache hit), is set to 0.86, which is the same as $\theta_{C}$ for clustering.

To measure the accuracy of outputs, we calculate the cosine similarity between the hit answer and its original answer (from the test dataset).
As shown in \FIG{fig:centvsopt}(a), \textsf{Centroid} achieves a cosine similarity of 0.85, while \textsf{Optimal} yields a higher similarity of 0.92, as it retains all of the vectors, using 10$\times$ more memory than \textsf{Centroid}.
Nevertheless, \textsf{Centroid}'s cosine similarity remains sufficiently high, showing its strong ability to capture relevant relationships.
The cosine similarity of \textsf{GPTCache} is higher than that of \textsf{Centroid}, since caching individual vectors occasionally retains vectors with cosine similarities close to 1.0, inflating the average.

\textbf{Impact on memory efficiency.}
To understand the impact of centroid caching on memory efficiency, we explore the hit ratio over the same three systems.
As shown in~\FIG{fig:centvsopt}(b), \textsf{Centroid} performs worse 
than \textsf{Optimal}, which has unlimited memory, but exhibits 1.27$\times$ and 1.14$\times$ higher hit ratios than \textsf{GPTCache} for Quora and Reddit, respectively.
This demonstrates the superior memory efficiency of \textsf{Centroid} over \textsf{GPTCache}.
To achieve comparable hit ratios, \textsf{GPTCache} requires 1.85$\times$ and 1.89$\times$ more memory for each dataset.

\lesson{Caching centroids, rather than individual vectors, provides high accuracy while being substantially more memory-efficient.}
\label{obs:observation2}

\subsection{Semantic Locality-aware Replacement}
\label{sec:semwarerp}

In the previous section, we assumed having a small yet sufficient memory to store all the centroids identified from datasets.
In practice, the semantic cache size is highly constrained, so centroids to cache must be carefully chosen among many candidates. Traditional strategies like LRU or LFU can be applied, but are inefficient.
Our analysis suggests that, for efficient centroid management, we must take into account another dimension of locality: \textit{semantic locality}~\cite{semantic-locality1,semantic-locality2}.

\textbf{Semantic locality.}
Embedding vectors of LLM queries are not uniformly distributed over the embedding space.
Instead, many similar vectors are highly localized or concentrated on certain regions, creating a few dense clusters, while other areas remain comparatively sparse.
Dense and sparse clusters can easily be identified through clustering.
Clusters that contain many similar vectors are considered dense and exhibit strong semantic locality; thus, their centroids must have high priority for caching, and vice versa.
As will be shown later, LRU or LFU, which exploit the recency or frequency of vectors, cannot capture such semantic relationships among vectors and thus fail to identify valuable vectors to cache.

To validate the above claim, we analyze the datasets collected from Quora and Reddit. 
We collect 60K centroids following the methodology described in~\SEC{sec:centbasedcaching}.
We then feed the queries from the test dataset to the system, counting the number of queries hit by the centroids using three different caching policies: \textsf{Semantic}, \textsf{LRU}, and \textsf{LFU}. 
The Semantic policy employs a static strategy: it initially fills the cache with centroids based on the number of queries they contain (a measure of semantic locality) until the cache limit is reached, without any subsequent replacements. 
In contrast, \textsf{LRU} and \textsf{LFU} are dynamic policies that promote centroids as they are accessed and evict a victim upon every cache miss.

\begin{figure}[t]
    \centering
        \begin{minipage}[t]{0.24\textwidth}
    	\centering
	        \includegraphics[origin=c,width=\linewidth]{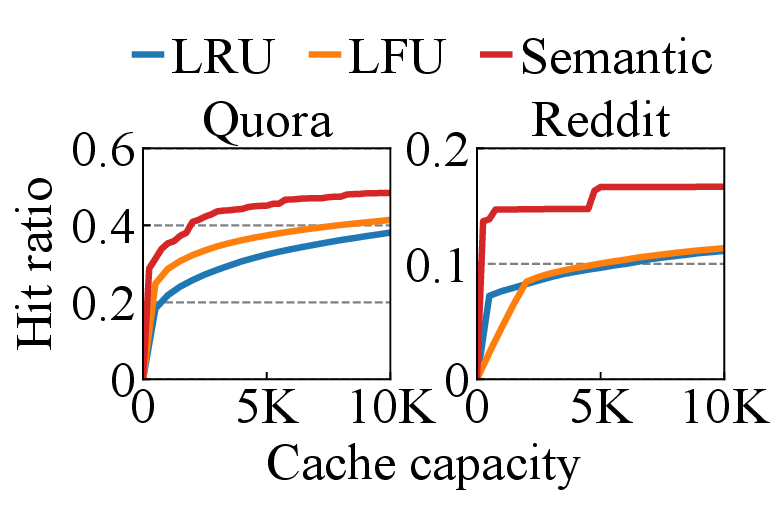}
            \caption{Hit ratios by policies}
	    \label{fig:locality}
        \end{minipage}
        \hspace{0.1pt}
        \begin{minipage}[t]{0.22\textwidth}
        \centering
            \includegraphics[origin=c,width=\linewidth]{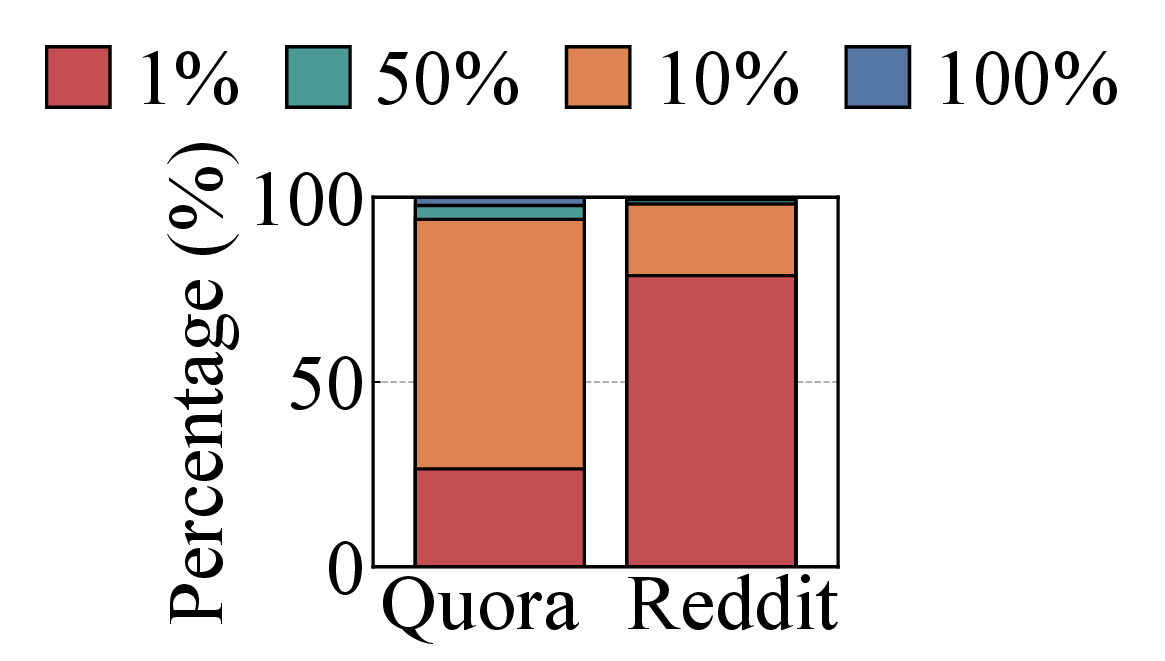}
            \caption{Ratio of rank change}
            \label{fig:rankstat}
        \end{minipage}
\end{figure} 

\FIG{fig:locality} presents the results, with the x-axis representing the cache capacity, defined as the number of cached centroids.
As shown, \textsf{Semantic} consistently outperforms both \textsf{LRU} and \textsf{LFU} in hit ratio across both datasets. 
This result is driven by two core distinctions: unlike heuristic policies, \textsf{Semantic} leverages semantic locality and performs no replacements.
Together, these lead to the hypothesis that semantic popularity is largely stable over time:
since a high semantic locality means that it contains a large number of related queries, which is an indication of the long-term popularity of such a centroid, it is valuable to retain them without eviction.

\textbf{Centroid robustness according to popularity.} 
To validate the previously set hypothesis, we conduct an experiment measuring the temporal stability of centroid popularity. Specifically, we track the rank of each centroid over four weeks and calculate the fraction of centroids whose rank change remained within 1\%, 10\%, 50\%, and 100\% of the total. 

As shown in \FIG{fig:rankstat}, 26.5\% and 78.8\% of the centroids in Quora and Reddit show less than 1\% rank variation. Moreover, 96.1\% of centroids change their rank by no more than 10\%, and only a minority (fewer than 4\%) experience substantial shifts. These results confirm that semantic locality remains largely stable with only gradual fluctuations over time. In practice, if we cache 10\% of the total centroids, only 2.4\% would require replacement to maintain an ideal hit ratio. Thus, aggressive replacement is unnecessary and misaligned with the long-term stability of centroid popularity.

\lesson{Since semantic locality remains relatively stable over time, replacements should be triggered only occasionally by monitoring long-term query behaviors.}
\label{obs:observation3}

\subsection{Dynamic Threshold Adjustment}
\label{sec:adjsimths}

\ours{} adjusts the similarity threshold $\theta_{R}$ dynamically
depending on the intensity of workloads
to balance the trade-off between the quality of responses and computational
efficiency. This threshold adjustment is feasible because of the linear
relationship between input and output similarities.

\begin{figure}[t]
    \centering
    \begin{minipage}[b]{0.27\textwidth}
        \centering
        \subfloat[Quora]{\includegraphics[width=0.37\linewidth]{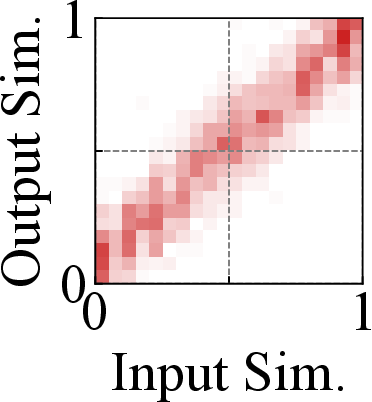}}
        \hspace{0.1pt}
        \subfloat[QQP]{\includegraphics[width=0.29\linewidth]{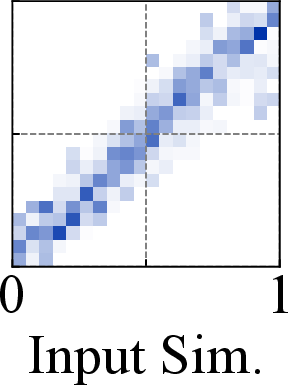}}
        \hspace{0.1pt}
        \subfloat[Reddit]{\includegraphics[width=0.29\linewidth]{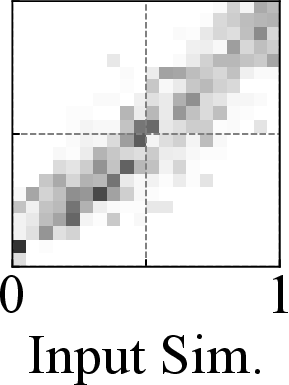}} \\
        \caption{Heatmap of input/output cosine similarities}%
        \label{fig:accesspattern}
    \end{minipage}
    \hspace{0.1pt}
    \begin{minipage}[b]{0.19\textwidth}
        \centering
        \subfloat[Quora]{\includegraphics[width=0.57\linewidth]{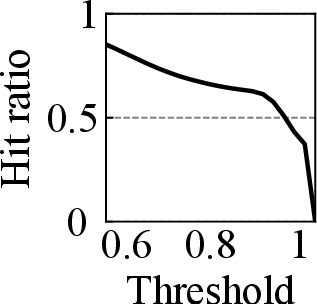}}
        \hspace{0.005\linewidth}
        \subfloat[Reddit]{\includegraphics[width=0.39\linewidth]{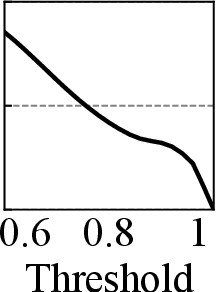}} \\
        \caption{Cache hit ratios with varying $\theta_R$}%
        \label{fig:adjst}%
    \end{minipage}
\end{figure}

To understand such a relationship, we analyze the
QQP, MRPC, MQP, Quora, and Reddit datasets we used
in~\SEC{sec:centbasedcaching} and~\SEC{sec:semwarerp}. 
For each dataset, we generate answers using the
LLaMa-3.1-8B model, creating pairs of <question, answer> as done previously.
We randomly choose two <question, answer> pairs, and then compute 
(\textit{i}) the input cosine similarity between the questions and 
(\textit{ii}) the output cosine similarity between the answers. 
\FIG{fig:accesspattern} shows the relationship between input and output
cosine similarities using a heatmap. 
The x- and y-axis represent the input and output similarities, respectively.
The color intensity in the heatmap corresponds to point density, with darker
regions indicating higher densities of <question, answer> pairs.
As depicted in \FIG{fig:accesspattern},
most data points align closely with the diagonal, showing a strong positive
correlation between input and output similarities.

\lesson{
There is a likelihood that semantically similar input queries (questions) will produce semantically similar outputs (answers).
}
\label{obs:observation4}

\textbf{Impact of $\theta_{R}$ on hit ratio.}
Observation \#4 indicates that the quality of outputs can be controlled
by adjusting $\theta_{R}$.
Setting $\theta_{R}$ high (\eg~0.98) ensures that only the centroids highly close
to input queries are selected, 
thereby producing outputs with high accuracy.
This, however, causes more frequent cache misses
due to a stricter similarity requirement, which in turn
increases the computational load on the LLM serving system.
Conversely, setting $\theta_{R}$ low (e.g., 0.60) 
increases the cache hit ratio, reducing the load on the LLM serving system.
However, this leads to a significant drop in the quality of responses due to
lower-similarity results.

\FIG{fig:adjst} shows cache hit ratios for $\theta_{R}$ varying 
from 0.60 to 0.99, using Quora and Reddit described in~\SEC{sec:centbasedcaching}.
At $\theta_{R}$ = 0.98, the cache hit ratio drops to 0.24, 
meaning that 76\% of input queries are
sent to the LLM serving system. In contrast, lowering $\theta_{R}$ to 0.60
increases the hit ratio to 0.85, reducing the LLM query traffic to 15\%. These results 
highlight the critical role of $\theta_{R}$ in balancing performance and accuracy: 
higher thresholds preserve response quality but increase load on the LLM,
whereas lower thresholds have the opposite effect.

\lesson{The similarity threshold is directly correlated to the output quality, but is inversely correlated to the hit ratio; thus it must be managed carefully considering the trade-off between output quality and latency.}
\label{obs:observation5}

\section{Implementation of \ours{}}
\label{sec:implementation}

\begin{figure}[t]
    \centering
    \includegraphics[width=\linewidth]{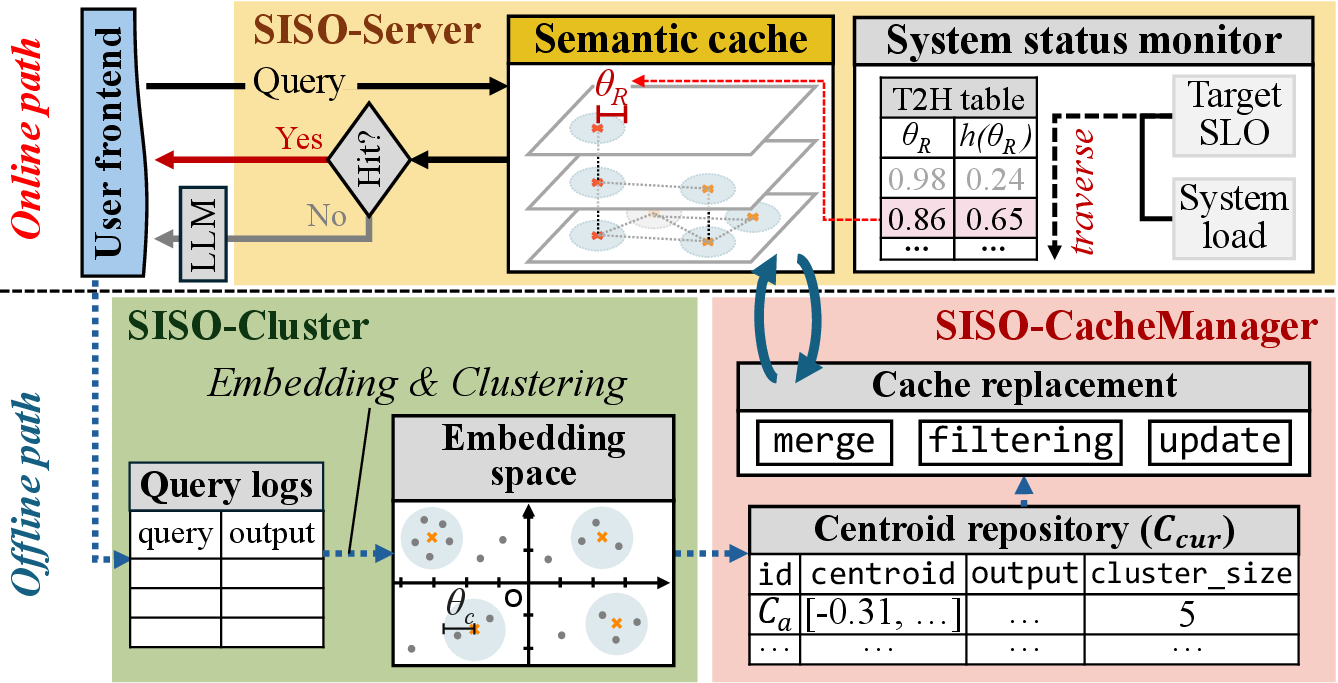}
    \caption{Implementation overview of \ours{}}
    \Description{Implementation overview of \ours{}}
    \label{fig:implementation}
\end{figure}

We now describe the implementation details of \ours{} shown in~\FIG{fig:implementation},
focusing on how we design \ours{} based on the key observations from~\SEC {sec:desprin}. 
\ours{} adopts a decoupled architecture with two paths,
an \textit{online path} for latency-intensive query serving and an \textit{offline path} for compute-intensive background tasks, ensuring that background tasks do not interfere with online query serving. 
The offline path consists of \textit{\ours{}-Cluster} for query clustering and \textit{\ours{}-CacheManager} for semantic cache management, while the online path is handled by \textit{\ours{}-Server}, which serves incoming queries in real time, adjusting the retrieval threshold $\theta_{R}$ according to workloads.
\SEC{sec:sisocluster}--\SEC{sec:sisoserver} detail these three components.

\subsection{Clustering: From Queries to Centroids}
\label{sec:sisocluster}
\ours{}-Cluster is responsible for extracting centroids 
from historical query logs by converting queries into embedding vectors,
clustering them, and choosing a representative centroid 
for each cluster. A selected centroid is stored in the centroid repository, along with
associated metadata (e.g., a generated output and the number of queries it represents). These centroids in the repository are considered candidates for promotion to the semantic cache.

To find high-quality centroids, it is preferable to use powerful embedding and clustering 
models. However, generating embedding vectors and forming clusters are compute-intensive tasks.
Moreover, those tasks must be repeated regularly 
as new queries accumulate. To avoid interfering with online serving,
\ours{}-Cluster should run on isolated machines, which requires additional
computing resources and thus raises operational costs. 
Therefore, it is crucial to choose embedding and clustering models that
can balance cost efficiency with centroid quality.

\begin{table}[t]
\centering
\caption{Candidate embedding models}
\label{tab:embmod}
\fontsize{7.5}{9.5}\selectfont
\begin{tabular}{l|cccc}
\hline
\multirow{2}{*}{\textbf{Model name }} & \multicolumn{2}{c|}{\textbf{Median cosine sim.}} & \multicolumn{1}{c|}{\multirow{2}{*}{\textbf{Gap}}} & \textbf{Latency} \\ 
\cline{2-3}
 & \multicolumn{1}{c|}{\textbf{Dup.}} & \multicolumn{1}{c|}{\textbf{Non-dup.}} & \multicolumn{1}{c|}{} & \textbf{(ms)} \\ 
\hline
all-mpnet-base-v2 & 0.59 & 0.89 & 0.30 & 4.94 \\
multi-qa-distilbert-cos-v1 & 0.59 & 0.88 & 0.29 & 3.27 \\
all-distilroberta-v1 & 0.58 & 0.86 & 0.28 & 2.96 \\
\textbf{paraphrase-albert-small-v2} & \textbf{0.59} & \textbf{0.86} & \textbf{0.27} & \textbf{2.63} \\
\hline
\end{tabular}
\end{table}

\textbf{Selecting an embedding model.}
To identify a suitable embedding model, 
we consider four candidates~\cite{all-mpnet-base-v2, multi-qa-distilbert-cos-v1, all-distilroberta-v1, paraphrase-albert-small-v2} in~\TAB{tab:embmod}.
Our goal is to find one that 
achieves short computation times with high accuracy in distinguishing duplicate queries.
Using the QQP dataset, we measure (\textit{i}) the median cosine similarities of vectors extracted from duplicate and non-duplicate pairs and (\textit{ii}) per-query processing time on a CPU.
The experiments are conducted in the same environment as in \SEC{sec:centbasedcaching}.
Based on the results, we choose \texttt{paraphrase-albert-small-v2}; it not only exhibits the shortest latency, but offers a sufficiently large similarity gap (i.e., 0.27) between duplicate and non-duplicate pairs, which means it identifies duplicate queries efficiently.

\textbf{Selecting a clustering algorithm.}
To cluster the extracted vectors, we consider four algorithms~\cite{dbscan, hdbscan, opticcs, comdet} in \TAB{tab:clustercomp}.
Since the number of clusters is unknown in our case, we exclude algorithms that require a specific cluster number as an input (e.g., K-means~\cite{kmeans}).
Using the QQP dataset with $\theta_{C}$ of 0.86, we measure the clustering time, the average and minimum cosine similarity between vectors in the same cluster.
The average similarity reflects the overall similarity of embedding vectors in the same cluster (higher is better). Conversely, 
the minimum cosine similarity tells us the worst-case quality of generated vectors 
in the cluster (higher is better).
OPTICS and HDBSCAN are excluded due to their prohibitively long runtimes, and DBSCAN, although faster, produces low minimum similarity, leading to noisy clusters.
We therefore choose Community Detection that achieves the highest minimum/average similarity with the shortest execution time.

\textbf{Keeping centroids in a repository.}
The clustering algorithm returns a unique identifier \texttt{id} of each
cluster, along with its centroid, \texttt{centroid}. For
each cluster, we maintain two additional elements, \texttt{output} and 
\texttt{cluster\_size}, in the repository.
\texttt{output} is the output text for the centroid. 
We use the output text of the input query that
is closest to the centroid.
\texttt{cluster\_size} is the number of vectors in the cluster,
which represents the degree of semantic locality. 
It is later used as a metric when choosing centroids to cache.

\begin{table}[t]
\centering
\caption{Comparison of various clustering algorithms}
\fontsize{7.5}{9.5}\selectfont
\begin{tabular}{l|ccc}
\hline
\textbf{Model name} &
  \textbf{Time (s)} &
  \textbf{\begin{tabular}[c]{@{}c@{}}Minimum\\ cos sim.\end{tabular}} &
  \textbf{\begin{tabular}[c]{@{}c@{}}Average\\ cos sim.\end{tabular}} \\ \hline
\textbf{Community Detection} & \textbf{41.44} & \textbf{0.80} & \textbf{0.99} \\
OPTICS                       & 34876.32       & 0.48          & 0.95          \\
DBSCAN                       & 82.15          & 0.39          & 0.97          \\
HDBSCAN                      & 29937.20       & -0.21         & 0.60\\
\hline
\end{tabular}%
\label{tab:clustercomp}
\end{table}

\textbf{Re-clustering.}
When and how often to trigger clustering is also a crucial issue due to its computational overhead. Initially, \ours{}-Cluster 
generates centroids by analyzing a long-term history of LLM queries (e.g., one year).
Medium-sized serving systems handle about 1B queries annually~\cite{cbot, burstgpt}.
Embedding and clustering 1B queries take about three days
on an AWS p3.2xlarge instance with one V100 GPU,
costing under \$300~\cite{comdetgpu}, which is reasonable, 
considering the long timespan.
Once the initial centroids are established, \ours{}-Cluster performs re-clustering
periodically to reflect the shifting semantics of new queries.
As shown in \OBS{obs:observation3}, this process does not need to be performed frequently.
In our evaluation (\SEC{sec:evaluation}), we trigger re-clustering when the number of newly accumulated queries reaches about 10\% of the initial query set
(it can be adjusted depending on workload characteristics and resource availability).
This approach enables us to amortize the clustering overhead while tracking changes in the semantic popularity of input queries.

\subsection{Caching: Centroids Replacement}
\label{sec:cachemanager}

Once the repository is updated,
\ours{}-CacheManager initiates cache replacement to
refresh the semantic cache. It compares existing centroids in
the semantic cache with newly discovered 
ones in the repository and decides which 
to insert, evict, or maintain.
To ensure uninterrupted query serving, the semantic cache should remain
active during replacement.

The operational flow of the cache replacement is detailed in
Algorithm~\ref{alg:recluster}, 
which is composed of three steps: merge, filtering, and update.
In the merge step, \ours{}-CacheManager combines
the set of centroids in the semantic cache, $C_{cur}$, with the set of new
centroids, $C_{repo}$, in the centroid repository 
and returns the newly combined set, $C_{new}$ (Line 1). In the filtering
step, it filters out centroids with low semantic locality or
with less popularity from $C_{new}$ (Line 2). 
In the update step,
it replaces $C_{cur}$ with $C_{new}$,
making the semantic cache up-to-date with new centroids (Line 3).

\begin{algorithm}[t]
\caption{\ours{}-CacheManager algorithm}
\label{alg:recluster}
\fontsize{8}{9.5}\selectfont
\Input{Current centroids $C_{cur}$, New centroids $C_{repo}$, Maximum memory capacity $M$, Threshold $\theta_C$}
\Output{Up-to-date current centroids $C_{cur}$}

$C_{new} \gets \MergeCentroids(C_{cur}, C_{repo}, \theta_C)$\;
$C_{new} \gets \FilteringCentroids(C_{new}, M)$\;
$\UpdateCentroids(C_{cur}, C_{new})$\;

\Return $C_{cur}$\;

\Fn{MergeCentroids($C_{cur}, C_{repo}, \theta_C$)}{
    $C_{new} \gets C_{cur}$\;
    \ForEach{$c_{repo} \in C_{repo}$}{
        $c_{closest} \gets \FindClosestCentroid(c_{repo}, C_{new})$\;
        \If{$\CosineSimilarity(c_{closest}, c_{repo}) > \theta_C$}{
            $c_{closest}[cluster\_size] \gets c_{closest}[cluster\_size] + c_{repo}[cluster\_size]$\;
        }\Else{
            $c_{repo}[access\_count] \gets \infty$\;
            $C_{new}.\text{add}(c_{repo})$\;
        }
    }
    \Return $C_{new}$\;
}

\Fn{FilteringCentroids($C_{new}, M$)}{
    \While{$\TotalMemoryUsage(C_{new}) > M$}{
        Sort $C_{new}$ by ($cluster\_size$, $access\_count$) in \scriptsize{ASC order}\;
        \footnotesize{Remove the first element from $C_{new}$\;}
    }
    \ForEach{$c \in C_{new}$}{
        $c[cluster\_size] \gets {c[cluster\_size]}/{1.1}$\;
        $c[access\_count] \gets 0$\;
    }
    \Return $C_{new}$\;
}

\end{algorithm}

\textbf{Merge step.} 
\ours{}-CacheManager first creates $C_{new}$ by copying current centroids from
$C_{cur}$ to $C_{new}$ (Line 6).  For each centroid $c_{repo}$ in $C_{repo}$,
it searches for the closest centroid 
$c_{closest}$ in $C_{new}$ based on cosine similarity (Line 8).
If the similarity between the
centroids exceeds the clustering threshold, $\theta_{C}$,
it means that the two represent the identical cluster.
Thus, we increase the cluster size, \textit{cluster\_size}, 
of $c_{closest}$ by adding that of $c_{repo}$ (Lines 9--10).
Recall that \textit{cluster\_size} is the number of vectors represented by a centroid, reflecting semantic locality.
If the similarity is below $\theta_{C}$, $c_{repo}$ is treated as a new centroid and 
thus added to $C_{new}$. 
Each centroid has an access count field, \textit{access\_count}, to
count how many times the centroid is referenced while in the semantic cache.
It reflects the short-term popularity of 
centroids.
Initially, \textit{access\_count} of
$c_{repo}$ is set to $\infty$ (Lines 12--13), which is to prioritize 
a new centroid $c_{repo}$ over old ones.  Finally,
\ours{}-CacheManager returns $C_{new}$ (Line 14).
During the merge step, \ours{} utilizes both metrics,
\textit{cluster\_size} and \textit{access\_count}, 
to ensure balanced query retention, effectively capturing 
both historical significance and recent popularity fluctuations.

\textbf{Filtering step.}
If the resulting centroid set, $C_{new}$,
exceeds the cache capacity,
\ours{}-CacheManager performs the filtering step to remove low-priority
centroids from $C_{new}$ (Line 16).
As discussed in~\SEC{sec:semwarerp}, the semantic locality
is a key criterion for deciding 
which centroids to evict.
\ours{}-CacheManager thus finds and removes the 
centroid with the smallest \textit{cluster\_size} from $C_{new}$.
If multiple centroids have the same cluster size, 
then we consider the popularity of centroids, \textit{access\_count}. That is,
the one with the smallest \textit{access\_count} is removed (Lines 17-18).
After resolving memory constraints, 
\ours{}-CacheManager scales \textit{cluster\_size} of $c_{new}$ down by 10\%
(Line 20).  
This reduction causes centroids that receive fewer new queries over time to gradually 
shrink in size, making them likely to be replaced.
Additionally, the access counts of all centroids are reset to zero
(Line 21), and the updated centroid set, $C_{new}$, is then returned as the
output (Line 22).

\textbf{Update step.} 
Finally, \ours{}-CacheManager begins replacing the old centroids, $c_{cur}$ in $C_{cur}$, with the new centroids, $c_{new}$ in $C_{new}$. Replacing all the centroids at once can cause serious locking overhead, negatively impacting the online path. 
To keep semantic caching available during updates, \ours{}-CacheManager progressively replaces small groups of centroids one at a time.
This avoids long blocking periods on the serving path and maintains consistent cache behavior throughout the replacement process.

\subsection{Serving: SLO-aware Query Serving}
\label{sec:sisoserver}

\ours{}-Server is in charge of serving input queries. 
While incoming queries are placed into a FIFO queue, 
\ours{}-Server fetches a batch of queries from the queue
and searches the semantic cache to identify similar queries.
Then, only the queries that aren't present in the cache are sent to the LLM.
To satisfy the target SLO, \ours{}-Server also 
adjusts $\theta_R$ to trade output quality for performance.
Note that users dissatisfied with cached responses may 
resubmit a similar query repeatedly.
To resolve this, \ours{} detects repeated queries from the same user and routes them directly to the LLM.

\textbf{Searching for vectors in semantic cache.}
Like other semantic caching systems, \ours{}-Server uses the
Hierarchical Navigable Small World (HNSW)~\cite{hnsw} algorithm to retrieve
similar vectors from the cache. To enhance search performance, we optimize HNSW
to exploit semantic locality.
In its standard form, HNSW randomly arranges vectors into a hierarchy of levels: upper levels contain only a small subset of vectors, while lower levels hold an increasingly larger collection. As the search moves downward, it visits progressively larger sets of vectors, enabling early termination of the search when a match is found in higher levels.

\ours{}-Server takes advantage of this hierarchical structure
by placing centroids with strong semantic
locality in the higher levels. Since these centroids are likely
to serve the majority of input queries, positioning them higher 
enables \ours{}-Server to quickly locate similar vectors earlier,
without descending to lower levels.

\textbf{Adjusting $\theta_{R}$ to meet SLO.}
To find the most suitable $\theta_{R}$, \ours{}-Server estimates the
average waiting time $W$ of the LLM system using the queuing theory. If $W$ is
longer than the desired SLO latency, 
we reduce the value
of $\theta_R$ to shorten $W$ with increased hit ratios, and vice versa.

To estimate the expected waiting time using the queuing theory,
we make three reasonable assumptions: (1) requests arrive following a Poisson process, which is a
common model for random arrival times of requests. (2)
serving time is stable with small fluctuation and thus deterministic.
(3) requests are processed in a FIFO manner.
Based on these assumptions, we model the LLM serving system as an
$M/D/1$ queue, where $M$ denotes a Poisson arrival process, 
$D$ represents deterministic service times, and $1$ indicates a single 
server.
In the M/D/1 model, $W$ can be expressed as follows:
\begin{equation}
W = E + \frac{\lambda E^2}{2(1-\lambda E)},
\end{equation}
where $E$ is the query serving time and $\lambda$ 
denotes the arrival rate of queries. In our system, $E$ is 
decided by two components: the latencies of semantic cache (on a cache hit) and the LLM
serving system (on a cache miss). 
The latency of the semantic cache is assumed to be zero
because it is much shorter than that of the LLM system.
Thus, $E$ can be defined as $E = L(1 - h(\theta_R))$,
where $L$ is the average serving time of the LLM system
and $h(\theta_R)$ is a cache hit ratio, which varies by $\theta_{R}$.
Consequently, the average waiting time $W$ can be rewritten as follows:
\begin{equation}
W = L(1-h(\theta_R))+\frac{\lambda L^2 (1-h(\theta_R))^2}{2(1-\lambda L (1-h(\theta_R)))}.
\label{eq:wait}
\end{equation}

In~\EQ{eq:wait}, $L$ is measurable from the LLM system
and $\lambda$ can be obtained by monitoring the arrival times of input queries.
Currently, we update $\lambda$ every ten seconds.
Our objective is to find the highest $\theta_R$ (or
the lowest $h(\theta_R)$) that satisfies $S > W$, where $S$ is
the desired SLO latency. 

Unfortunately, $h(\theta_R)$ is unknown and varies depending
on workloads. To estimate the expected hit ratio $h(\theta_R)$ by $\theta_{R}$, 
\ours{} constructs a threshold-to-hit-ratio (T2H) table that maps $\theta_R$ to a corresponding hit ratio $h(\theta_R)$. 
The T2H table is built during the clustering process;
once the clustering finishes and the semantic cache is updated, 
\ours{} samples
5\% of the newly arrived queries from the query log at random. 
Then, while varying $\theta_R$ from 0.98 to 0.6, 
\ours{} performs test lookups on the semantic cache using sample queries and 
measures hit ratios.
Constructing the T2H table typically takes only a few minutes,
and thus its overhead is minimal. 

\begin{table}[t]
\centering
\footnotesize
\caption{Information of datasets}
\label{tab:dataset_info}
\begin{tabular}{l|cc|cc} 
\hline
\textbf{Dataset} & \begin{tabular}[c]{@{}c@{}}\textbf{\# of}\\\textbf{queries}\end{tabular} & \begin{tabular}[c]{@{}c@{}}\textbf{Avg. \# of}\\\textbf{tokens}\end{tabular} & \begin{tabular}[c]{@{}c@{}}\textbf{\% of simple}\\\textbf{queries}\end{tabular} & \begin{tabular}[c]{@{}c@{}}\textbf{\% of complex}\\\textbf{queries}\end{tabular} \\ 
\hline
MSMARCO & 1M & 7 & 95.1\% & 4.9\% \\
NQ & 310K & 9 & 95.9\% & 4.1\% \\
Quora & 600K & 12 & 93.1\% & 6.9\% \\
Reddit & 631K & 14 & 56.9\% & 43.1\% \\
ShareGPT & 92K & 112 & 53.4\% & 46.6\% \\
\hline
\end{tabular}
\end{table}

Using the T2H table, \ours{}-Server can promptly find the most suitable
$\theta_R$ at runtime. However, 
the estimated $W$ could differ significantly from the actual $W^{'}$.
If the error exceeds 10\%, \ours{}-Server adjusts $\theta_R$ further.
For example, if $W{'}$ is much longer than $W$, \ours{}-Server lowers $\theta_R$ 
to reduce the query traffic to the LLM system.

\section{Evaluation}
\label{sec:evaluation}

\subsection{Experimental Setup}
\label{sec:experimental-setup}

We evaluate \ours{} on a GPU server
equipped with an Intel Core i5-8600K
CPU, 1.5TB DRAM, 28TB NVMe SSD, and eight 48GB A6000 GPUs.  
vLLM is used as a baseline LLM serving system.  
We implement \ours{}
from scratch and run it as a separate module in the same machine
with vLLM. vLLM utilizes GPUs for inference,
while \ours{} uses CPUs and DRAM. 
We use two LLM models~\cite{llama3}: a smaller LLaMa-3.1-8B, and a larger LLaMa-3.1-70B.
Following a common practice~\cite{aladdin}, the target SLO latency is set to 1.3$\times$ the E2E serving latency, which is defined as 
\( TTFT 
\: + TBT  \: \times ( \# \: of \: generated \: tokens -1) 
\).

We use various real-world datasets to demonstrate a broad applicability of \ours{} (see~\TAB{tab:dataset_info} for detail).
Besides the Quora and Reddit datasets used throughout~\SEC{sec:centbasedcaching}--~\SEC{sec:adjsimths}, we include MSMARCO~\cite{msmarco} and Natural Questions (NQ)~\cite{natural_questions} with relatively short input tokens, as well as ShareGPT~\cite{sharegpt} with much longer input queries. By employing datasets with diverse token lengths and query characteristics, such as simple (\eg~information seeking) and complex queries (\eg~coding), we ensure that our results are not confined to specific dataset types. 
For more information, please refer to Appendix~\ref{sec:category-detail}.
Unless otherwise stated, we utilize 95\% of each dataset for training and the remaining 5\% for testing. 

We configure vLLM so that the lengths of outputs for queries follow the
distribution of outputs' lengths observed in ShareGPT. 
This is for fair comparison across systems: 
longer generations incur higher latency in vLLM, whereas cached responses are returned instantly regardless of length. Aligning the output length distribution prevents bias in favor of caching systems.
To evaluate various workloads with different variability, 
we simulate request arrivals using a Poisson process, 
adjusting both the average arrival rate and the coefficient of variation (CV). 
We also simulate the intensity of workloads 
by varying requests-per-second (RPS).

\begin{figure*}[t]
    \centering
    \begin{minipage}{0.33\textwidth}
    \centering
        \includegraphics[width=0.99\linewidth]{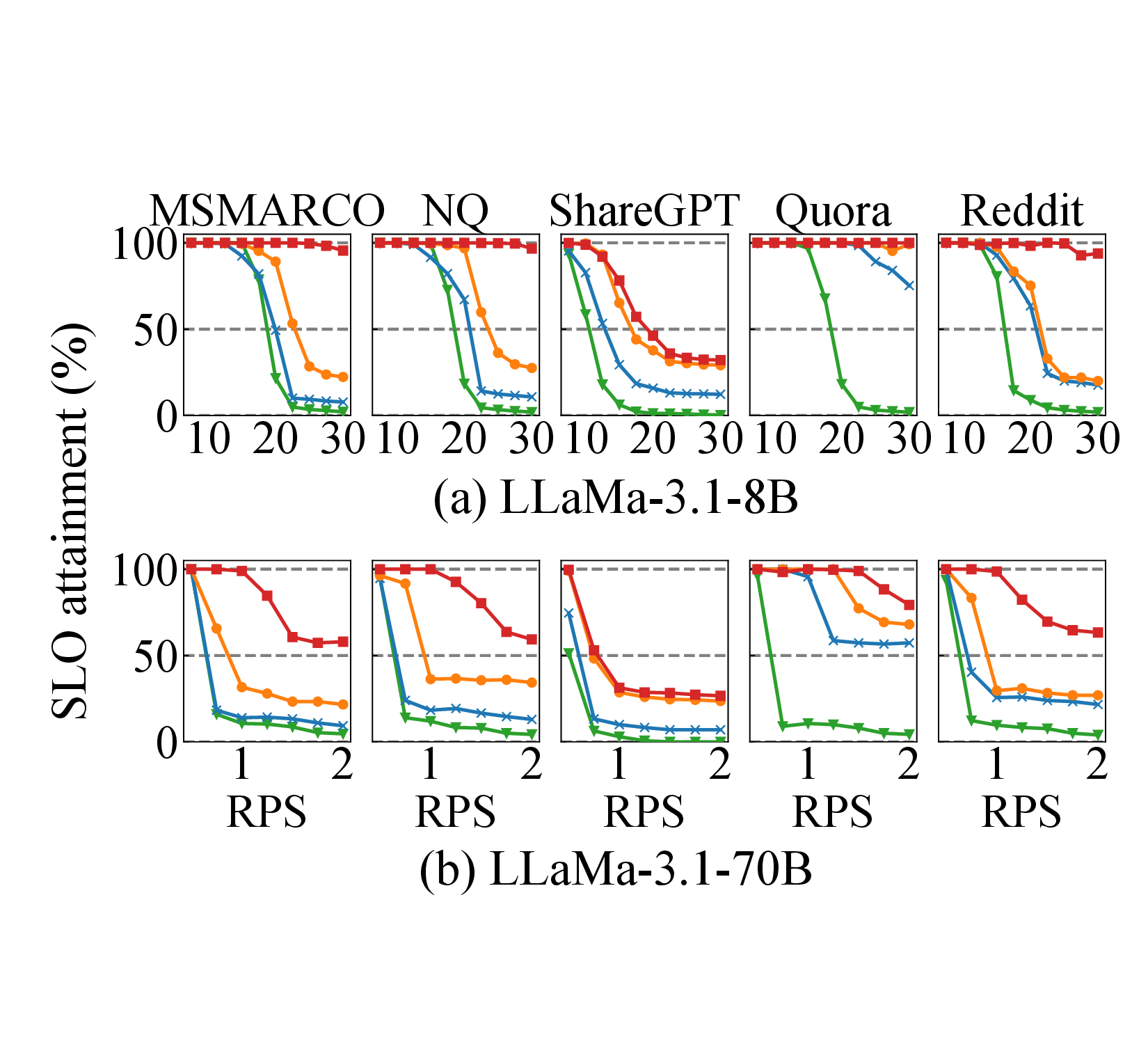}
        \caption{Impact of RPS on SLO attainment}
        \label{fig:slo-atn}
    \end{minipage}
    \begin{minipage}{0.33\textwidth}
    \centering
        \includegraphics[width=0.99\linewidth]{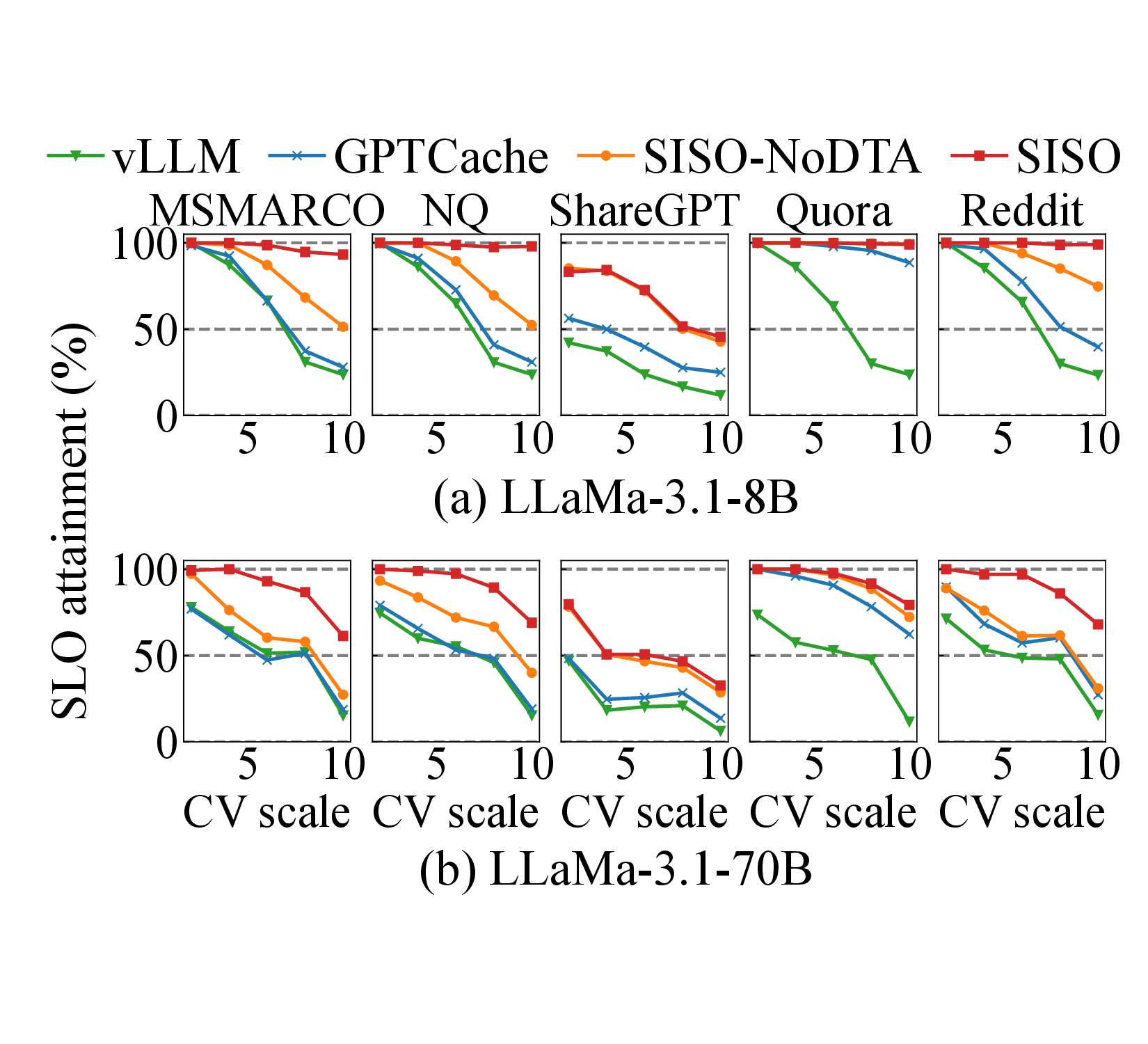}
        \caption{Impact of CV on SLO attainment}
        \label{fig:cv_scale}
    \end{minipage}
    \begin{minipage}{0.33\textwidth}
	\centering
	\includegraphics[width=0.99\linewidth]{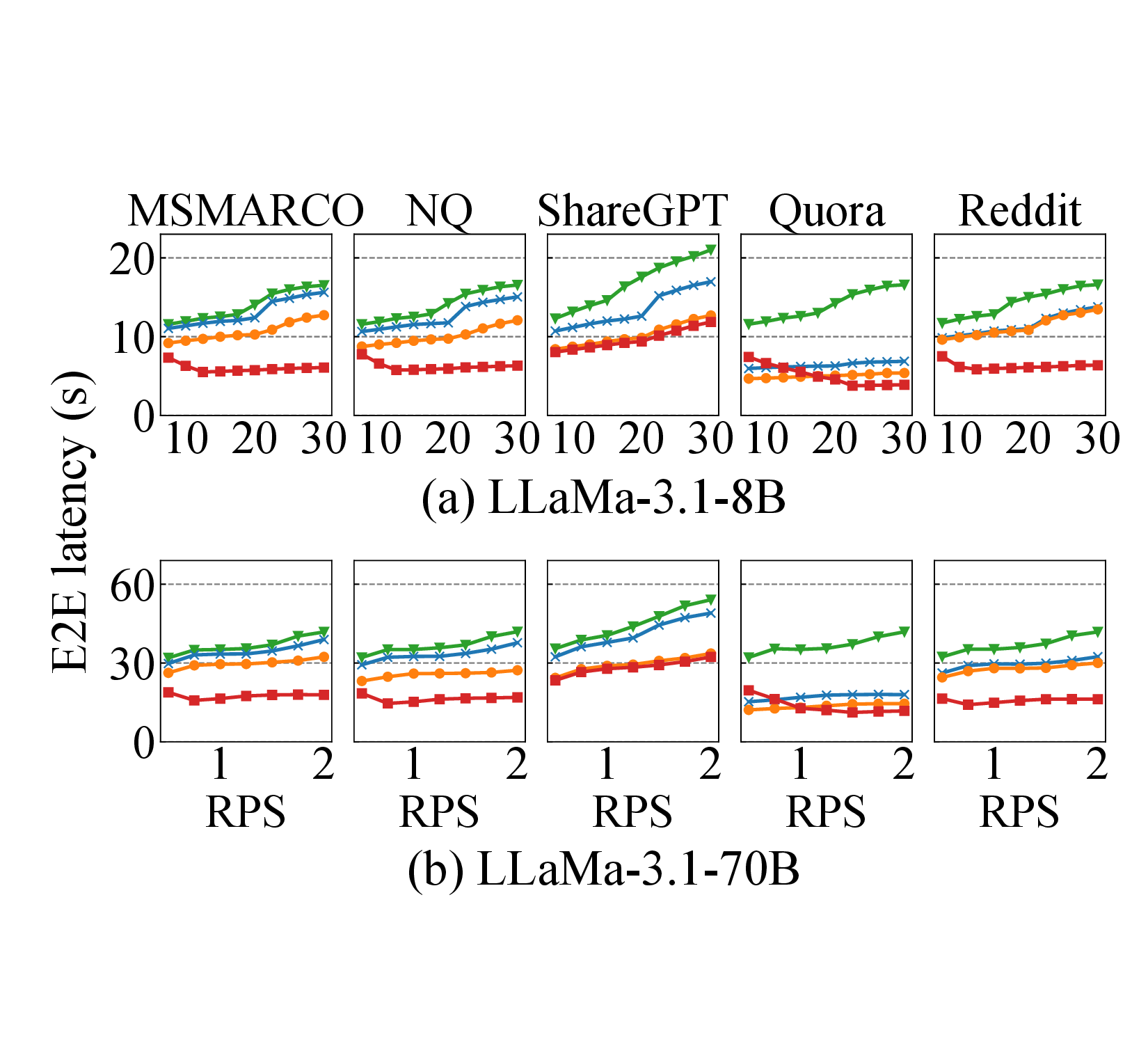}
	\caption{Impact of RPS on E2E latency}
        \Description{Impact of RPS on E2E latency}%
	\label{fig:atgt}%
    \end{minipage}
\end{figure*}

Each dataset contains a different number of queries, 
making it impractical to evaluate all systems with a fixed capacity. 
For this reason,
we scale the semantic cache size in proportion to the dataset size. Since
there are no specific guidelines for cache size, we set the cache
capacity to accommodate 6\% of the embedding vectors from each dataset.
This number is conservatively chosen by referring to
practical LLM deployments. Large-scale services
such as
ChatGPT are estimated to have processed over 38TB of query logs~\cite{chatgpt-stat}.
6\% of such queries would require approximately 2.3TB of memory -- an amount well within the capability of high-end inference clusters or vector database systems~\cite{aws_ec2, nvidia_dgx_h100, googlecloud_m2}.
We also perform evaluation while varying the semantic cache size.

We compare four different systems: \textsf{vLLM}, \textsf{GPTCache},
\ours{} without dynamic threshold adjustment, denoted by
\textsf{\ours{}-NoDTA}, and \textsf{\ours{}}. \textsf{vLLM} is the LLM serving
system that does not employ any semantic caching. \textsf{GPTCache} is
the SOTA semantic caching system that manages individual embedding vectors using LRU.  
It uses vLLM as an underlying LLM serving system.  We include
\textsf{\ours{}-NoDTA} to understand the impact of dynamic threshold adjustment
on performance. \textsf{\ours{}} employs all the techniques we explained
in~\SEC{sec:desprin} and~\SEC{sec:implementation}.  For both \textsf{GPTCache} and
\textsf{\ours{}-NoDTA}, $\theta_{R}$ is fixed to 0.86, while \textsf{\ours{}}
dynamically adjusts it.

\subsection{Experimental Results}
In this subsection, we present results on SLO attainments under
varying workload intensity (\SEC{sec:slo-atn}) and variability (\SEC{sec:cv-scale}),
analysis of serving latency (\SEC{sec:atgt}--\SEC{sec:analysis-llm-latency}), 
impact of cache size on hit ratio (\SEC{sec:hit-ratio-vs-cache-size}), and response quality
(\SEC{sec:response-quality}).

\subsubsection{Impact of Workload Intensity on SLO}
\label{sec:slo-atn}

We first evaluate the SLO attainment rate of the four systems as RPS increases.
We set the CV to a low value, 0.1, 
to minimize the variability in request patterns.
The results are shown in \FIG{fig:slo-atn}.
In MSMARCO and NQ,  \textsf{\ours{}-NoDTA} exhibits
higher SLO attainment rates than \textsf{GPTCache}, highlighting 
the effectiveness of the centroid-based caching over LRU.
However, at approximately 20 RPS,
its SLO attainment declines due to the rising query volume. In contrast, 
\textsf{\ours{}} maintains a stable and high attainment even under heavy load --
satisfying the target SLO at RPS close to 30 -- by dynamically adjusting $\theta_{R}$
to increase the likelihood of cache hits and reduce the number of queries sent to the LLMs.

In Quora, \textsf{\ours-NoDTA} and \textsf{GPTCache} show
high SLO attainment rates as well.  This is due to high semantic cache
hit ratios that reach 50\%.  In contrast, in ShareGPT, the SLO
attainment sharply drops as RPS increases for all the systems. 
This stems from ShareGPT queries containing longer input tokens than other workloads (see~\TAB{tab:dataset_info}),
which increases KV cache usage during inference and consequently reduces the number of requests that can be processed per second.

For LLaMa-3.1-70B with the same GPU configuration, 
the SLO attainment rate drops at lower RPS compared to the smaller model. 
This is because the larger LLaMa-3.1-70B model
significantly increases resource demands for LLM inference, 
thereby limiting the number of requests it can handle at the same RPS. 
Despite this, \textsf{\ours{}} achieves better SLO attainment rates than other systems.
This demonstrates the effectiveness of \textsf{\ours{}} even for large-scale models.
As the trends observed in~\FIG{fig:slo-atn} align with results from other experiments,
we omit detailed explanations for LLaMa-3.1-70B unless otherwise noted.

\subsubsection{Impact of Workload Variability on SLO}
\label{sec:cv-scale}

We perform experiments while varying CV from 2 to 10 with a fixed RPS of 8 for LLaMa-3.1-8B and 0.5 for LLaMa-3.1-70B, respectively. 
A higher CV implies more irregular request patterns, whereas
a lower CV indicates stable request patterns.

As illustrated on Fig.~\ref{fig:cv_scale}, higher
CV causes short-term spikes in workload patterns
that increase computational demand, thereby reducing overall SLO attainment.
All systems except for \textsf{\ours{}}
fail to meet the SLO under higher CV conditions. 
\textsf{\ours{}} successfully maintains high SLO attainment rates 
across all datasets other than ShareGPT, 
which indicates that \textsf{\ours{}} can
capture sudden changes in input workloads and 
effectively adjust $\theta_{R}$ to meet SLO. 
In ShareGPT, the improvement on the cache hit ratio by lowering
$\theta_{R}$ is limited to 1.36$\times$, much lower than
the other workloads that show 12.6$\times$ improvements, on average
(see~\SEC{sec:discussion} for more explanations).

\subsubsection{Impact of Workload Intensity on E2E Latency}
\label{sec:atgt}
Fig.~\ref{fig:atgt} presents the changes in E2E serving latency
as RPS increases.
As expected, \textsf{\ours{}} exhibits the lowest latency
regardless of RPS.
Interestingly, at very low RPS, \textsf{\ours{}} exhibits higher latency.
Particularly, in Quora that exhibits strong
semantic locality, \textsf{\ours{}} shows longer latency than
\textsf{\ours{}-NoDTA}. 
Under light workloads,
\textsf{\ours{}} decides to prioritize the quality of responses by increasing $\theta_{R}$, sending more input queries to the LLM. 
This decision is reasonable -- if the LLM
serving system has enough capacity to deal with input queries
without violating the SLO, improving the output quality is a preferable strategy.

\subsubsection{Analysis of Serving Latency} 
\label{sec:analysis-llm-latency}

\TAB{tab:time-breakdown} shows the breakdown of the average E2E serving latency
with LLaMa-3.1-8B for \textsf{\ours{}} and \textsf{GPTCache}. 
When a request is hit by the cache, 
\textsf{\ours{}} and \textsf{GPTCache} provide very short response times,
17ms and 27ms, respectively, since heavy LLM computation
that takes about 12 seconds is skipped.
By using optimized HNSW where centroids with strong semantic locality are placed
in higher levels and thus less traversal is needed,
\textsf{\ours{}} was able to achieve 
1.7$\times$ faster cache lookup performance than \textsf{GPTCache} on cache hit.
Although semantic cache lookup adds an extra overhead in serving
queries when a cache miss happens, it is negligible compared to the
computational cost of the LLM and has minimal impact on LLM serving latency.

\begin{table}
\centering
\footnotesize
\caption{Breakdown of serving latency}
\label{tab:time-breakdown}
\begin{tabular}{l|c|cc|cc} 
\hline
 & \multicolumn{1}{l|}{\multirow{2}{*}{\textbf{vLLM}}} & \multicolumn{2}{c|}{\textbf{GPTCache }} & \multicolumn{2}{c}{\textbf{SISO }} \\ 
\cline{3-6}
 & \multicolumn{1}{l|}{} & Hit & Miss & Hit & Miss \\ 
\hline
\textbf{Embedding (ms)} & - & 2.63 & 2.63 & 2.63 & 2.63 \\
\textbf{Search (ms)} & - & 23.98 & 23.99 & 13.92 & 16.16 \\
\textbf{Inference (s)} & 11.98 & - & 11.91 & - & 11.89 \\ 
\hline
\textbf{Total (s)} & 11.98 & 0.027 & 11.94 & 0.017 & 11.91 \\
\hline
\end{tabular}
\end{table}

\subsubsection{Analysis of Hit Ratio with Varying Cache Size} 
\label{sec:hit-ratio-vs-cache-size}

\begin{figure}[t]
	\centering
	\includegraphics[width=0.98\linewidth]{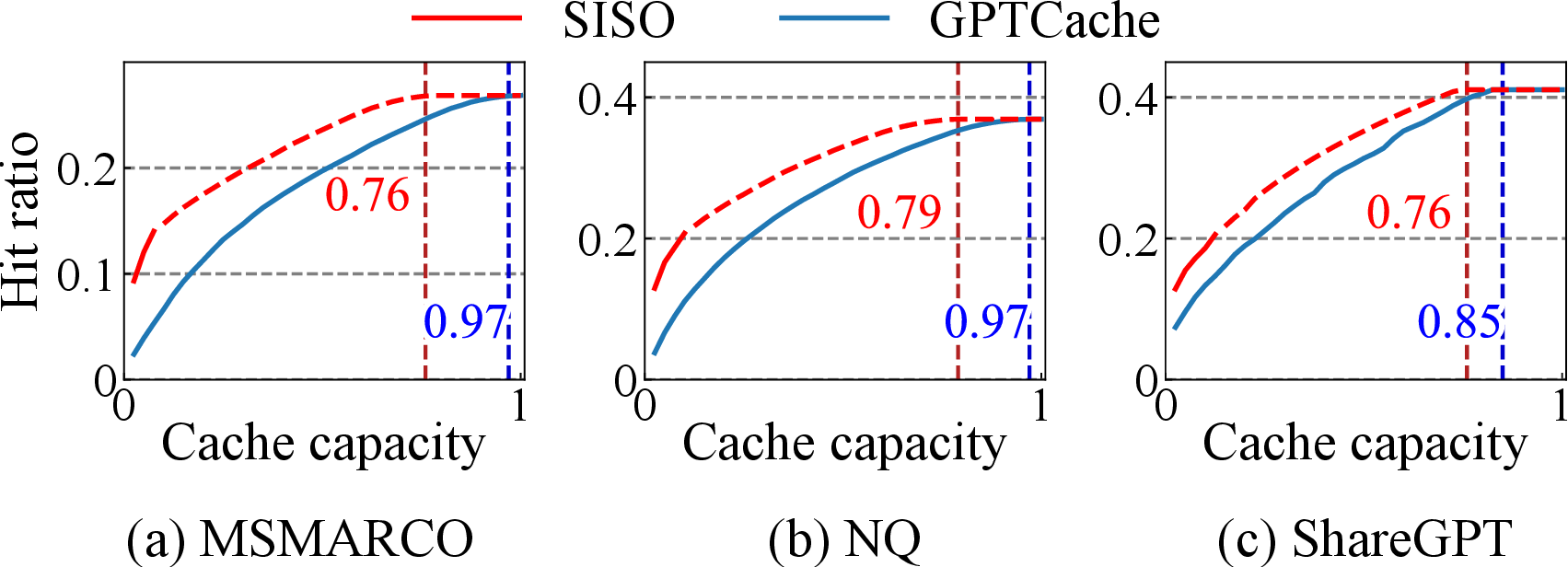}
	\caption{Hit ratios by cache capacity}%
        \Description{Hit ratios by cache capacity}%
	\label{fig:hit-ratio}
\end{figure}

\FIG{fig:hit-ratio} compares the cache hit ratios of \textsf{\ours{}} and
\textsf{GPTCache} depending on cache size with $\theta_{R}$ fixed at 0.86.
\textsf{\ours{}} stores only centroid vectors in the cache, but
if free space remains,
\textsf{\ours{}} caches individual vectors and manages them using LRU.
Since the results for Quora and Reddit are already presented in ~\FIG{fig:locality}, 
we only present MSMARCO, NQ, and ShareGPT.

Under constrained cache sizes, 
\textsf{\ours{}} achieves a higher hit ratio
than \textsf{GPTCache}.  For example, in MSMARCO, \textsf{\ours{}}
provides the hit ratio of 15\% with the cache capacity of 10\%, 
which can only be
accomplished by \textsf{GPTCache} when its cache capacity is
3$\times$ that of \textsf{\ours{}}. 
This is since \ours{} can identify clusters 
with strong semantic locality and cache small-sized centroids.  
In contrast, \textsf{GPTCache} ignores
semantic relationships of vectors,
which leads to premature eviction of
important vectors.
As cache capacity increases, both systems eventually reach their maximum achievable hit ratios.
However, \textsf{\ours{}} attains this peak earlier than \textsf{GPTCache}, thus requiring significantly less memory.

\begin{figure}[t]
	\centering
	\includegraphics[width=0.99\linewidth]{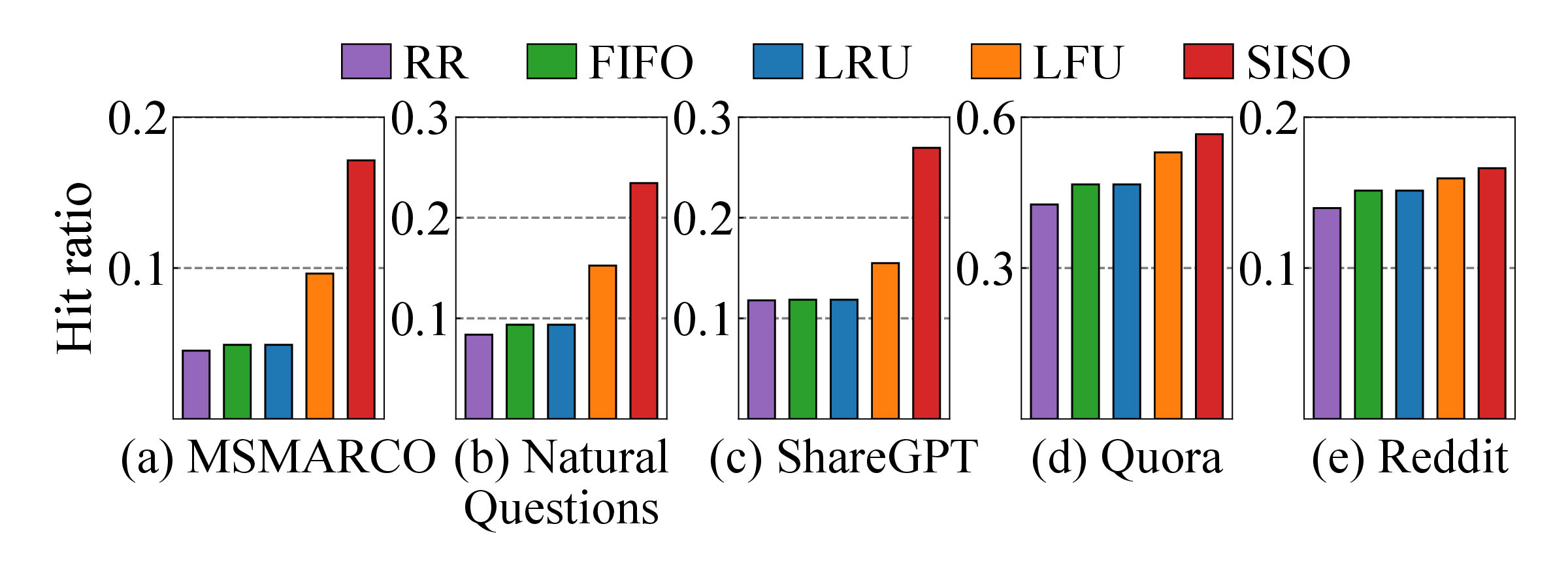}
	\caption{Comparison of various cache replacement policies} 
        \Description{Comparison of various cache replacement policies} 
	\label{fig:GPTCache-policy-siso}%
\end{figure}

\subsubsection{Analysis of Various Replacement Policies} 
\label{sec:various-replacement-policy}
We evaluate the impact of various cache replacement policies on cache hit
ratios. We implement round-robin (RR), FIFO, and LFU policies in \textsf{GPTCache}
and compare their hit ratios with \textsf{\ours{}}. Note that \textsf{GPTCache} employs
LRU as its default policy. The cache capacity is set to 6\% of the dataset size.

~\FIG{fig:GPTCache-policy-siso} demonstrates that centroid-based caching 
outperforms the other replacement policies, achieving 
43\% improvement on hit ratios
than the next best policy, LFU, on average. These results confirm our claim:
selecting centroids through the clustering process based on the long-term history 
of LLM serving is more effective than relying solely on the recency, frequency, or
sequence of input queries.

\subsubsection{Analysis of Response Quality}
\label{sec:response-quality}

\begin{figure}[t]
    \centering
    \includegraphics[width=0.999\linewidth]{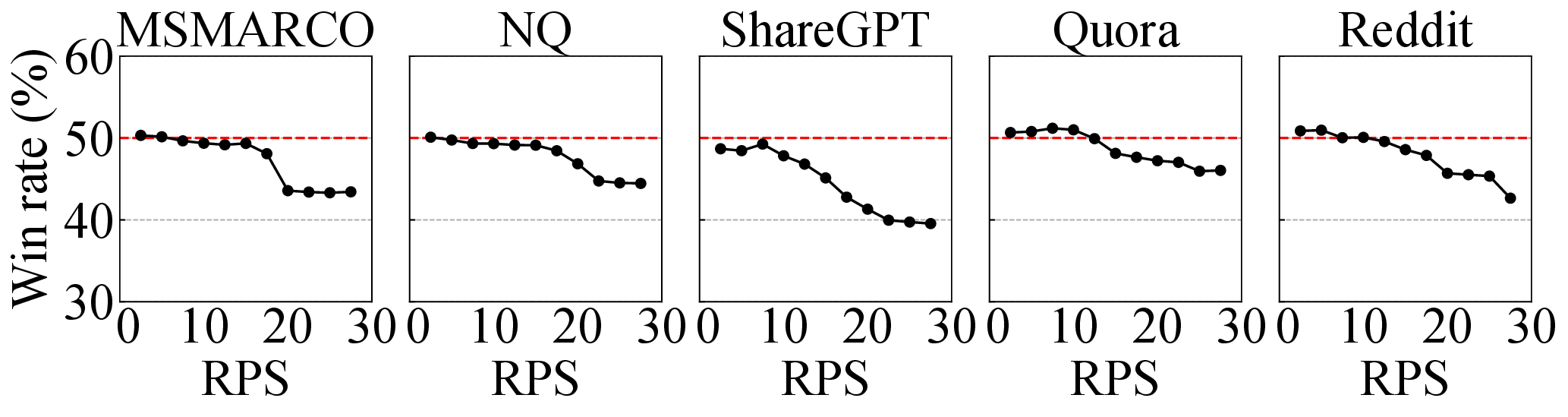}
    \caption{Win rate of \textsf{\ours{}} against \textsf{vLLM}}
    \label{fig:alpaca_eval_results}
\end{figure}

We now evaluate the quality of outputs from \textsf{\ours{}}.
For our assessment, we use Alpaca-eval~\cite{alpacaeval},
which is a widely used benchmark tool 
for LLM response quality evaluation.
It uses GPT-4 to compare the quality of outputs between a target model and a reference model: we use \textsf{vLLM} as 
the reference and \textsf{\ours{}} as the target, both based on LLaMa-3.1-8B.

\FIG{fig:alpaca_eval_results} illustrates the Win rates of \textsf{\ours{}}
under varying RPS, where a Win rate indicates 
the percentage of cases 
where \textsf{\ours{}}'s outputs are judged superior. A Win rate near 50\% 
indicates that \textsf{\ours{}} produces high-quality outputs comparable 
to \textsf{vLLM}.
At low RPS, \textsf{\ours{}} shows Win rate close to 50\%.
As RPS increases, the Win rate of \textsf{\ours{}} begins to drop
as it trades accuracy for performance. However, even at a high RPS close to 30,
\textsf{\ours{}} maintains sufficiently high Win rates -- 42.2\% -- on average.
This implies that \textsf{\ours{}} can gain efficiency
with minimal impact on response quality.

\begin{figure}[t]
	\centering
	\includegraphics[width=0.99\linewidth]{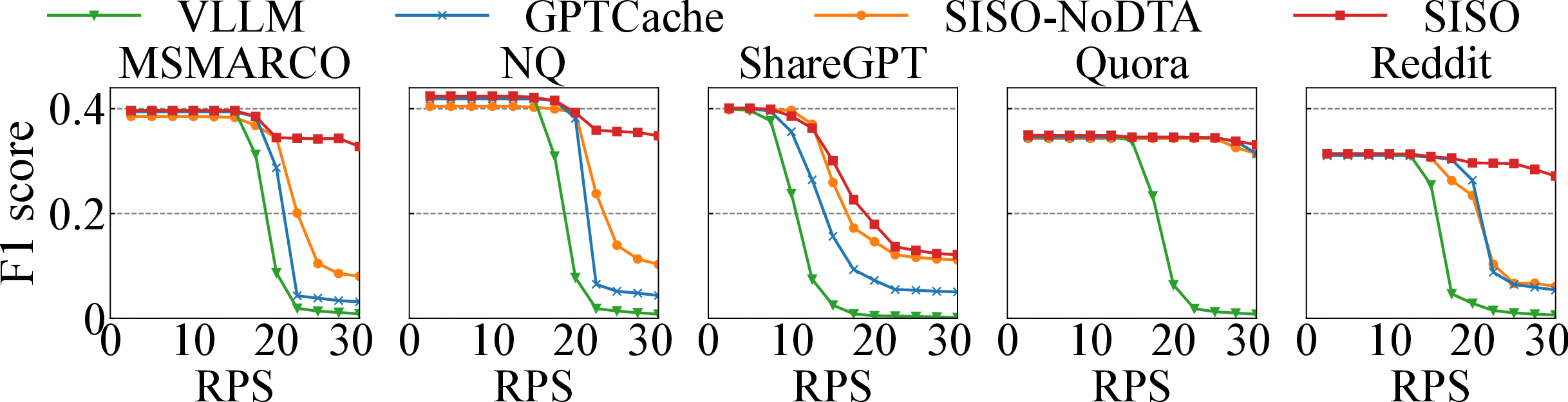}
	\caption{Comparison of workload F1 scores by RPS}
	\label{fig:f1score}%
\end{figure}

To quantitatively evaluate the response quality, we calculate F1 scores
by comparing each system's output against GPT-4 responses, which serve as reference answers. Unlike Win rates, which are a preference-based measure, the F1 score directly quantifies output quality by counting token-level overlaps.
To reflect user‑perceived quality, SLO‑violating requests are scored as 0, 
as they receive HTTP 429 (``Too Many Requests'') without any meaningful responses~\cite{mooncake}.

\FIG{fig:f1score} shows F1 scores by dataset and RPS. At low RPS, 
\textsf{\ours{}} performs comparably to the other techniques with no or negligible F1 score drops. 
As RPS increases, \textsf{\ours{}} gradually lowers $\theta_{R}$ to reduce requests to the LLM and suppress SLO violations,
leading to a higher F1 score than other systems.
On average, \textsf{\ours{}} achieves 1.17x, 1.26x, and 1.71x higher F1 score than \textsf{SISO‑NoDTA}, \textsf{GPTCache}, and \textsf{vLLM}, respectively.

 \section{Discussion and Limitation}
\label{sec:discussion}

\ours{} assumes that similar inputs yield similar outputs.
This assumption, however,
does not hold if the
output relies heavily on the context, which is the case for multi-turn queries.
For instance, ``What should I pick for tomorrow?'' 
could mean dinner menus vs. travel plans depending on the context.
Yet, the multi-turn queries are not dominant in either API calls (<1\%) or chatbots (33\%), so the impact on \ours{} is limited.
Still, addressing such cases remains an open challenge, which could be interesting future work.

Likewise, queries where minor input changes significantly alter the output (e.g., coding and debugging) limit the effectiveness of semantic caching.
This explains \ours{}'s lower performance on ShareGPT~(\SEC{sec:slo-atn}--\SEC{sec:atgt}), which contains such queries, including coding, debugging, and brainstorming.
To assess this deeper, we evaluate the SLO attainment with LLaMa-3.1-8B across four representative categories
under increasing RPS using the same systems and methodology 
as in \SEC{sec:slo-atn}.
For our analysis, we construct categorized datasets by extracting
queries from all the datasets used 
in~\SEC{sec:evaluation} and grouping them by category using the method in~\cite{query-class}.

\begin{figure}[t]
	\centering
	\includegraphics[width=0.9\linewidth]{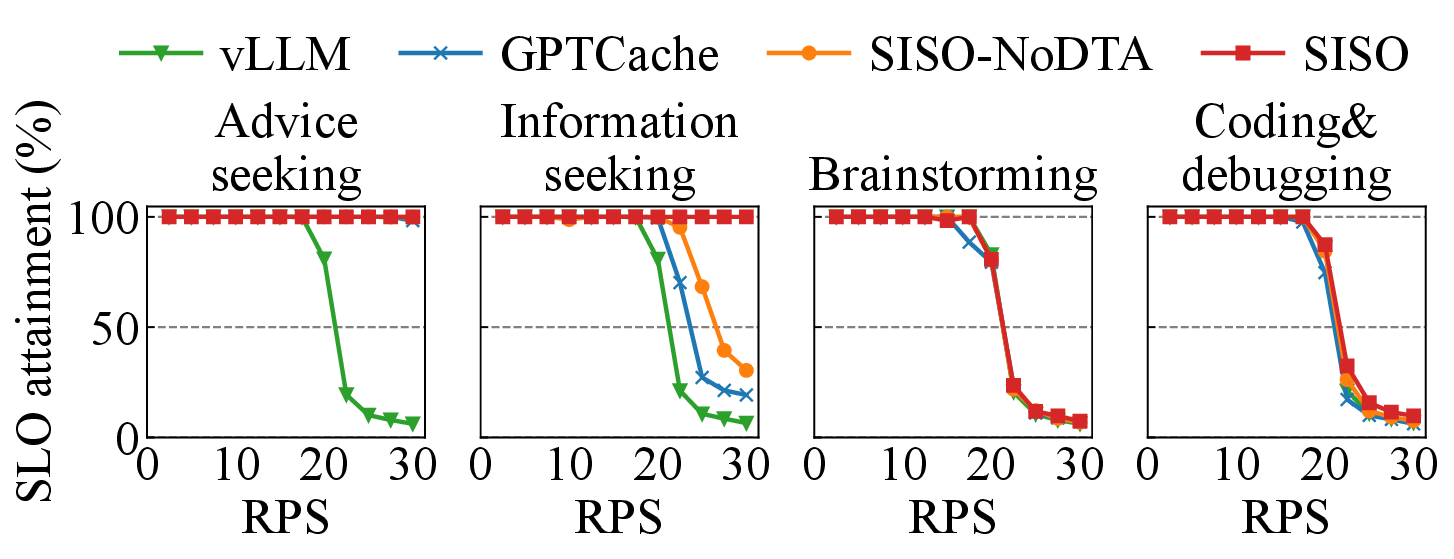}
	\caption{SLO attainment across categories}
	\label{fig:category-slo}%
\end{figure}

As shown in \FIG{fig:category-slo}, the observed outcomes align with the anticipated trends.
\ours{} 
excels in Advice and Information seeking categories, which are predominantly single-turn and thus yield high hit ratios.
Conversely, the performance gains are less pronounced for Brainstorming and Coding \& debugging, where small input changes result in significant differences in outputs.
Nevertheless, these complex categories account for only a minority of real-world workloads (code: 13.5\%, brainstorming: 5.2\%), whereas Q\&A accounts for the majority (53.1\%)~\cite{mossdataset}. 
Importantly, even under these challenging conditions, \ours{} consistently outperforms \textsf{vLLM} and \textsf{GPTCache}.
Taken together with results from \SEC{sec:slo-atn}--\SEC{sec:atgt}, this demonstrates that \ours{} delivers robust performance across datasets that reflect real-world query distributions.

\section{Conclusion}
\label{sec:conclusion}

In this paper, we presented \ours{}, an enhanced semantic caching system.
\ours{} employed centroid-based caching,
semantic locality-aware centroid replacement, and dynamic threshold adjustment, which
improved cache hit ratios and ensured consistent SLO compliance even during peak loads. 
Our experiments showed that \ours{} outperformed the SOTA systems, 
achieving 1.71x, on average, higher hit ratios and enhanced SLO attainment rates.
These improvements were obtained while maintaining high quality of responses 
comparable to the reference model, vLLM.

\clearpage
\bibliographystyle{ACM-Reference-Format}

\clearpage
\appendix
\section*{Appendices}

\section{Category Detail for Each Dataset}
\label{sec:category-detail}

\begin{figure}[h]
    \centering
    \includegraphics[width=0.9\linewidth]{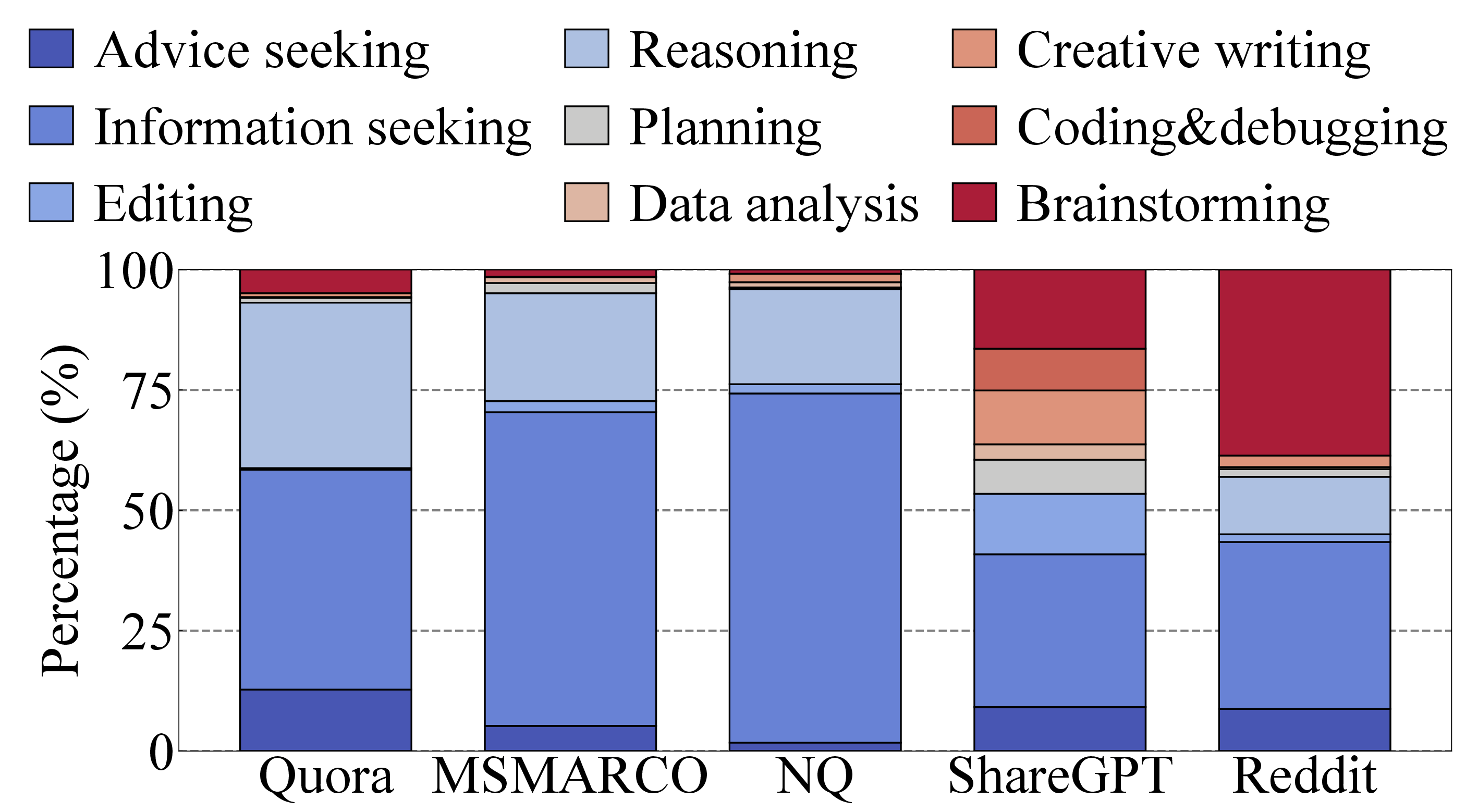}
    \caption{Distribution of query categories across dataset}
    \label{fig:category-detail}
\end{figure}

In \SEC{sec:discussion}, we measured performance across different categories. Overall,  \ours{} exhibited relatively poor performance on tasks such as Coding\&debugging and Brainstorming.
In addition, through our categorization method~\cite{query-class}, we identified a total of nine categories, and their distribution is shown in \FIG{fig:category-detail}. The topics in the shade of blue (Advice seeking, Information Seeking, Editing and Reasoning) are simple queries, and the ones in the shade of red (Planning, Data analysis, Creative writing, Coding\&debugging and Brainstorming) are complex queries. These also correspond to the categories shown in~\TAB{tab:dataset_info}.

In these categories, Editing and Reasoning showed experimental results similar to advice seeking and information seeking (higher performance gain) presented in \FIG{fig:category-slo}, while data analysis and creative writing exhibited trends comparable to Coding\&debugging and Brainstorming (lower performance gain). This characteristic arises, as noted earlier, in tasks where small variations in the input query can lead to significant differences in the output response. 


As we mentioned, such complex tasks constitute only a small fraction of the overall workload. In this sense, \ours{} still demonstrates promising performance from a general workload perspective.

\section{Hit Ratio Varying RPS}

In this section, we present the hit ratio as the RPS is varied.
As other experimental results also indicated, \ours{} drops the hit ratio on low demand to improve the output quality.
This drop only occurs when the LLM serving system has sufficient capacity for LLM execution, and it does not negatively impact the system.

\begin{figure}[h]
    \centering
    \subfloat[LLaMa-3.1-8B]{\includegraphics[width=0.99\linewidth]{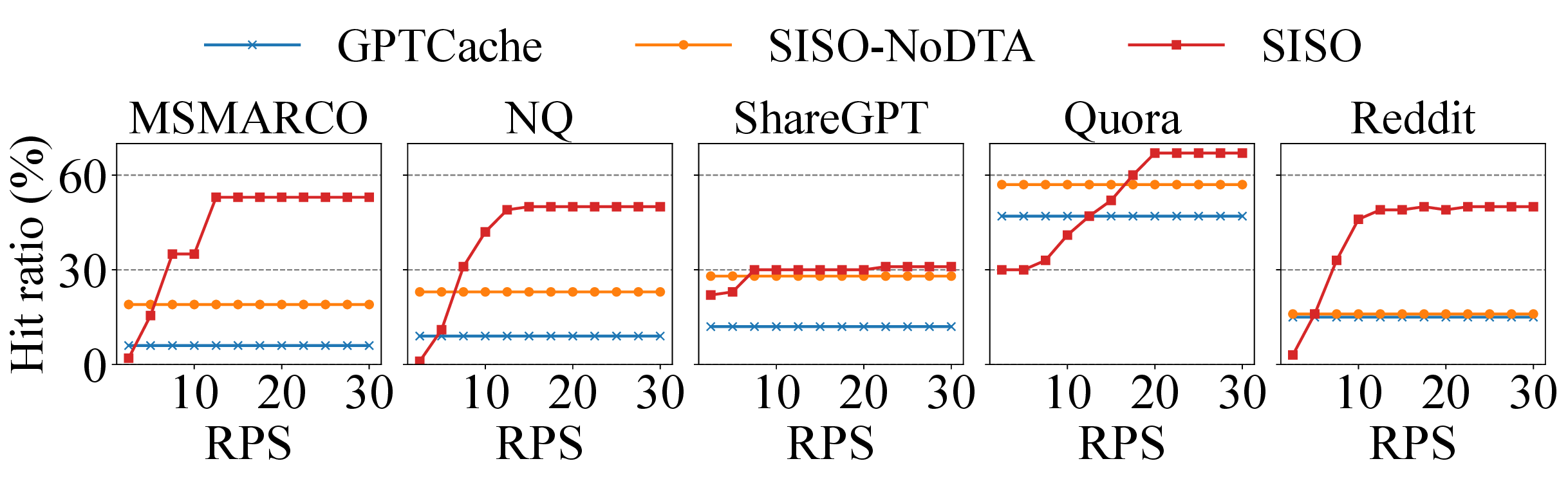}}\\
    \subfloat[LLaMa-3.1-70B]{\includegraphics[width=0.99\linewidth]{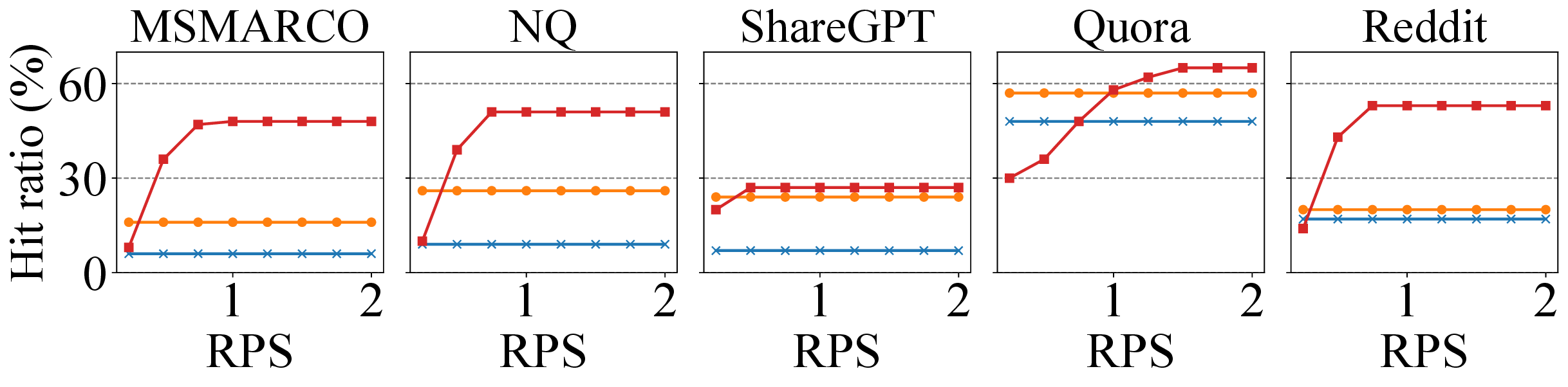}}
    \caption{Impact of RPS on hit ratio}
    \label{fig:rps-hitratio}%
\end{figure}

\end{document}